\documentclass[12pt]{article}
\usepackage{amsmath}
\usepackage{times}
\usepackage{graphicx}
\usepackage{color}
\usepackage{multirow}
\usepackage[authoryear]{natbib}
\usepackage{rotating}
\usepackage{bbm}
\usepackage{latexsym}
\usepackage[tablesfirst]{endfloat}

\newcommand{\captionfonts}{\normalsize}

\makeatletter  
\long\def\@makecaption#1#2{%
  \vskip\abovecaptionskip
  \sbox\@tempboxa{{\captionfonts #1: #2}}%
  \ifdim \wd\@tempboxa >\hsize
    {\captionfonts #1: #2\par}
  \else
    \hbox to\hsize{\hfil\box\@tempboxa\hfil}%
  \fi
  \vskip\belowcaptionskip}
\makeatother   



\begin{document}
\hspace{13.9cm}1

\ \vspace{20mm}\\

{\LARGE Inferring neuronal couplings from spiking data using \textcolor{black}{a} systematic procedure with a statistical criterion}

\ \\
{\bf \large Yu Terada}\\
{\textit{yu.terada@riken.jp}}\\
{Laboratory for Neural Computation and Adaptation, RIKEN Center for Brain Science, 2-1 Hirosawa, Wako, Saitama 351-0198, Japan}\\
{\bf \large Tomoyuki Obuchi}\\
{\textit{obuchi@i.kyoto-u.ac.jp}}\\
{Department of Systems Science, Graduate School of Informatics, Kyoto University, Kyoto 606-8501, Japan}\\
{\bf \large Takuya Isomura}\\
{\textit{takuya.isomura@riken.jp}}\\
{Laboratory for Neural Computation and Adaptation, RIKEN Center for Brain Science, 2-1 Hirosawa, Wako, Saitama 351-0198, Japan}\\
{\bf \large Yoshiyuki Kabashima}\\
{\textit{kaba@phys.s.u-tokyo.ac.jp}}\\
{Institute for Physics of Intelligence, Graduate School of Science, The University of Tokyo, Tokyo 113-0033, Japan}\\
%


\thispagestyle{empty}
\markboth{}{NC instructions}
\ \vspace{-0mm}\\
%
\begin{center} {\bf Abstract} \end{center}
Recent remarkable advances in the experimental techniques have provided a background for inferring neuronal couplings from point process data that includes a great \textcolor{black}{number} of neurons.
Here, we propose a systematic procedure for pre- and post-processing generic point process data in an objective manner, to handle data in the framework of a binary simple statistical model, the Ising or generalized McCulloch--Pitts model.
The procedure involves two steps:  
(1) determining time-bin size for transforming the point-process data into discrete-time binary data and (2) screening relevant couplings from the estimated couplings.
For the first step, we decide the optimal time-bin size by introducing the null hypothesis that all neurons would fire independently, then choosing a time-bin size so that the null hypothesis is rejected with the most strict criterion.
The likelihood associated with the null hypothesis is analytically evaluated and used for the rejection process.
For the second post-processing step, after a certain estimator of coupling is obtained based on the pre-processed dataset (any estimator can be used with the proposed procedure), the estimate is compared with many other estimates derived from datasets obtained by randomizing the original dataset in \textcolor{black}{the} time direction.
We accept the original estimate as relevant only if its absolute value is sufficiently larger than them of randomized datasets. 
These manipulations suppress false positive couplings induced by statistical noise. 
We apply this inference procedure to spiking data from synthetic and \textit{in vitro} neuronal networks. 
The results show that the proposed procedure identifies the presence/absence of synaptic couplings fairly well including their signs, for the synthetic and experimental data. 
In particular, the results support that we can infer the physical connections of underlying systems in favorable situations, even when using the simple statistical model.

\section{Introduction}

Recent tremendous technological advances have enabled us to measure multi-point signals simultaneously, which may reveal many natures of the central nervous system and other biological systems \citep{weigt2009identification,lezon2006using}.
\textcolor{black}{Multi-point neuronal activities} can be recorded by microelectrode arrays \textcolor{black}{\citep{buzsaki2004large,brown2004multiple,steinmetz2018challenges}} or calcium imaging \citep{ikegaya2004synfire,cheng2011simultaneous,grewe2010high}.
These types of \textcolor{black}{studies} provide rich data and help us \textcolor{black}{to} understand \textcolor{black}{the} mechanisms of information processing in large coupled systems beyond the single-neuron level \citep{yuste2015neuron,roudi2015multi,grosmark2016diversity,maass2016searching,paninski2018neural}.

Theoretical results from the field of the statistical physics have offered powerful tools \textcolor{black}{for inferring intrinsic structures} from such measurements.
In \textcolor{black}{cases} with \textcolor{black}{undirected} (symmetric) couplings\textcolor{black}{,} where the systems are in equilibrium, the mean-field formulae for statistical inference have been previously developed \citep{kappen1998efficient,tanaka1998mean,sessak2009small,roudi2009ising} and successfully applied to biological data \citep{schneidman2006weak,shlens2006structure,cocco2009neuronal,tang2008maximum,ohiorhenuan2010sparse}.
Recently, beyond undirected cases, more general cases, or those with directed (asymmetric) connectivity structures in nonequilibrium states can be handled with reasonable computation by using improved systematic techniques \citep{roudi2011mean,mezard2011exact,zeng2011network,sakellariou2012effect,aurell2012inverse,zeng2013maximum}, which instigates their applications for real data \citep{tyrcha2013effect,dunn2015correlations}.

\begin{figure}
  \centering
	\includegraphics[scale=0.28]{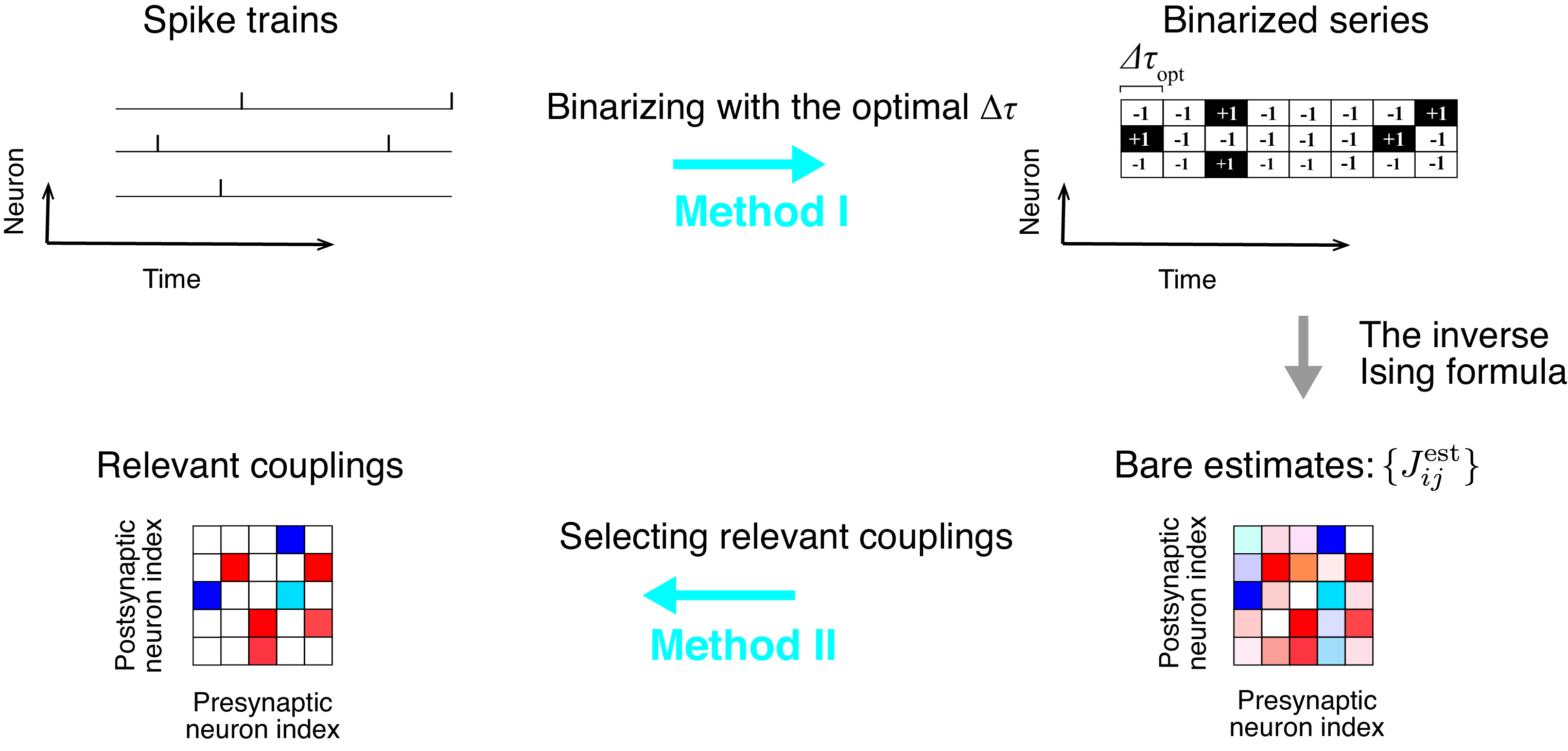}
  \caption{Entire procedure to reconstruct couplings.
  Using multi-point observations, we first obtain event-series data that records the times when events occurred.
  We use the criterion Eq. \eqref{eq:opt_tau} to make the original series binned with $\Delta\tau_{\text{opt}}$.
  Applying the inverse mean-field formula, we estimate the coupling matrix and obtain \textcolor{black}{the bare estimates $\{J_{ij}^{\text{est}}\}$}, whose elements are continuous.
  Finally, we select \textcolor{black}{relevant} couplings from them using a computational-statistical method with randomized sequences.}
  \label{fig:coarse_grain}
\end{figure}

While statistical-mechanics approaches have revealed the nontrivial nature of biological systems, there still remain unsatisfactory points in earlier studies. 
Two crucial points are the lack of an objective criterion to determine bin size for temporal discretization of signals and to effectively screen relevant couplings obtained by inference techniques.
In this paper, we describe an objective procedure to resolve these issues based on the methods of information theory (Method I in Fig. \ref{fig:coarse_grain}) and computational statistics (Method II in Fig. \ref{fig:coarse_grain}), respectively.
These methods can be applied to a wide variety of dynamical systems that exhibit event sequences irrespective of the directionality of connectivities. 
As a representative example, we apply these methods to a mathematical model of plausible neuronal networks called the Izhikevich model \citep{izhikevich2003simple}.
Consequently, we can reconstruct the underlying couplings of the networks with high accuracy, even when using the simple kinetic Ising model as the inference model.
Motivated by this finding, we also apply these methods to {\it in vitro} neuronal networks of rat cortical cells cultured in a simple circular structure \textcolor{black}{\citep{isomura2015signal}.
Thanks to this physical constraint, the anatomical connections of the cultured system are expected to also have the circular structure.
The inferred couplings derived by our inference methods actually exhibit the clear circular structure expected by the experimental design.}
The result suggests that the kinetic Ising model, despite its simplicity, can be a good, reliable estimator of the sign pattern of true synaptic couplings.

\section{Inference model and methods}

\subsection{Inference model --- the kinetic Ising model}

\textcolor{black}{In this work, we use} the kinetic Ising model\textcolor{black}{,} which consists of $N$ elements with \textcolor{black}{two} possible discrete state values\textcolor{black}{:} $s_i(t)=\pm1$.
The $s_i(t)=1$ \textcolor{black}{state} corresponds to the firing of the neuron $i$\textcolor{black}{, whereas the} $s_i(t)=-1$ \textcolor{black}{state} means no firing at time $t$.
This model is supposed to obey Galuber dynamics:
\begin{align}
P\left(\mathbf{s}(t+1)|\mathbf{s}(t)\right) = \prod_{i=1}^{N}\frac{\exp\left[s_i(t+1)H_i(t)\right]}{\exp\left[H_i(t)\right]+\exp\left[-H_i(t)\right]},
\end{align}
where $H_i(t)$ is the effective field defined as $H_i(t)=h_i(t)+\sum_{j=1}^{N}J_{ij}s_j(t)$, $h_i(t)$ is the external force, and $J_{ij}$ is the coupling strength from $j$ to $i$.
This model corresponds to the generalized McCulloch--Pitts model \citep{mcculloch1943logical} in theoretical neuroscience and logistic regression \citep{cox1958regression} in statistics.

\textcolor{black}{In this study, the maximum likelihood method forms the basis for inferring the parameters.
However, a direct maximization of likelihood is often accompanied with plenty of computational cost, which can be a bottleneck in post-processing of screening.
Therefore, we use a simple mean-field formula to reduce the computational cost.
This approximation keeps the inference accuracy in terms of coupling signs as shown in Sec. \ref{sec:goodness}.
The \textcolor{black}{direct} maximum likelihood method and other more sophisticated formulae can also be principally \textcolor{black}{applied} in combination with our methodology.
The mean-field inverse formula derived by Roudi and Hertz in \citep{roudi2011mean} is written as $J^{\text{est}}=A^{-1}DC^{-1}$, where $m_i=\left<s_i(t)\right>$, $C_{ij}=\left<s_i(t)s_j(t)\right>-m_im_j$, $D_{ij}=\left<s_i(t+1)s_j(t)\right>-m_im_j$ and $A_{ij}=\left(1-m_i^2\right)\delta_{ij}$. 
 The bracket represents the time average, which is expected to be accordance with the ensemble average under the ergodic assumption and we make this assumption in this work \citep{churchland2007techniques,cunningham2014dimensionality}. }

Throughout this paper, we assume that \textcolor{black}{the data do not involve dominant long-range collective modes of statistical fluctuations that interfere with the accurate coupling inference \citep{das2019systematic}.
This marks the limit of the application range of the proposed methods}.
\textcolor{black}{Although such modes are sometimes observed in some natural setups, data without the long-range modes \textcolor{black}{have recently attracted} much attention as they can be observed, for example, in the neuronal activity during tasks \citep{dahmen2019second,stringer2019high} as well as the spontaneous activity.
Therefore, we focus on data of this type.
Note that the absence of the long-range modes in data can be tested by a statistical method as shown in Sec. \ref{sec:limitation}.}

\subsection{Selecting optimal bin size}

For the kinetic Ising model and other types of generalized linear models \citep{okatan2005analyzing,pillow2008spatio,field2010functional,kobayashi2019reconstructing}, we have to determine the size of time bins when pre-processing raw data.
Although it is known that the size of the time bins in binarizing the spiking data can affect the inference results \citep{capone2015inferring}, objective criteria for the bin size have not been considered sufficiently.
To address this issue, we employ an information-theoretic method to determine the bin size in an objective way.

To apply the inverse kinetic Ising scheme to spike train data, we firstly determine the length $\Delta\tau$ of the time bin from an information theory viewpoint \citep{kullback1997information}\textcolor{black}{, in the procedure shown in} Fig. \ref{fig:coarse_grain}.
We begin with the spike series\textcolor{black}{,} which is obtained with a measurement frequency $\Delta t^{-1}$\textcolor{black}{, which} decides the minimal time unit (i.e. temporal resolution).
We divide the spike trains \textcolor{black}{using time bins of $\Delta\tau$}, binarize them, and obtain a spike raster with the whole time-length $T$, to ensure that each bin takes \textcolor{black}{a} +1 or -1 value to express the presence or absence of spikes in \textcolor{black}{a} bin, respectively.
Then, to operate the inverse formula effectively\textcolor{black}{,} their time bins are made coarse-grained with the appropriate $\Delta\tau$\textcolor{black}{, which} corresponds to the unit time in the kinetic Ising system.

In preparation \textcolor{black}{to calculate the optimal $\Delta\tau$ size}, we suppose that the system reached \textcolor{black}{its} stationary state, and set a null hypothesis that each neuron would fire or not fire independently of the other neurons.
We consider a method to determine the optimal bin size $\Delta\tau$ such that this hypothesis is rejected \textcolor{black}{with the most strong criterion}.
\textcolor{black}{For} each pair of neurons $i$ and $j$, we denote the numbers \textcolor{black}{how many} binarized states in successive bins $(s_i(t+1),s_j(t))$ fall on the four possible patterns $(+1,+1)$, $(+1,-1)$, $(-1,+1)$, and $(-1,-1)$ among $M-1(=T/\Delta \tau-1)$ time bins, as $n^{ij}_{++}$, $n^{ij}_{+-}$, $n^{ij}_{-+}$, and $n^{ij}_{--}$, respectively.
Under the null hypothesis,  \textcolor{black}{the neurons} $i$ and $j$ are considered to fire with probabilities of $p_{+}^{i\cdot}=1-p_{-}^{i\cdot}=(n_{++}^{ij}+n_{+-}^{ij})/(M-1)$ and $p_{+}^{\cdot j}=(n_{++}^{ij}+n_{-+}^{ij})/(M-1)$, respectively. 
Given the coarse-grained data with $n_{++}^{ij}, n_{+-}^{ij}, n_{-+}^{ij}, n_{--}^{ij}$, this provides us with the likelihood of the hypothesis as
\begin{align}
&\prod_{i\ne j}P\left(n_{++}^{ij},n_{+-}^{ij},n_{-+}^{ij},n_{--}^{ij}|p_{+}^{i\cdot},p_{+}^{\cdot j}\right) \notag\\
&=\prod_{i\ne j}\frac{(M-1)!}{n_{++}^{ij}!n_{+-}^{ij}!n_{-+}^{ij}!n_{--}^{ij}!}\left(p_{+}^{i\cdot}p_{+}^{\cdot j}\right)^{n_{++}^{ij}}\left(p_{+}^{i\cdot}(1-p_{+}^{\cdot j})\right)^{n_{+-}^{ij}}\left((1-p_{+}^{i\cdot})p_{+}^{\cdot j}\right)^{n_{-+}^{ij}}\notag\\
&\,\,\,\,\,\,\left((1-p_{+}^{i\cdot})(1-p_{+}^{\cdot j})\right)^{n_{--}^{ij}}\notag\\
&\simeq\exp\left[-(M-1)\sum_{i\ne j}I_{\Delta\tau}\left(s_i(t+1);s_j(t)\right)\right],\label{eq:aprrox_stirling}
\end{align}
where the last expression \textcolor{black}{in Eq.} \eqref{eq:aprrox_stirling} is derived by the Stirling's formula and
\begin{align}
I_{\Delta\tau}\left(s_i(t+1);s_j(t)\right) = \sum_{(\alpha,\beta) \in \{+,-\}^2} \frac{n^{ij}_{\alpha\beta}}{M-1}\log\frac{\frac{n^{ij}_{\alpha\beta}}{M-1}}{p^{i\cdot}_{\alpha}p^{\cdot j}_{\beta}}
\end{align}
represents the mutual information of neurons $i$ and $j$ between successive time bins.
Notation $\{+,-\}^2$ denotes \textcolor{black}{the} direct product $\{+,-\} \times \{+,-\}$. 
In Eq. \eqref{eq:aprrox_stirling}, contributions from self-interactions are omitted to focus on interactions between different neurons. 
Our aim is to determine the bin size $\Delta \tau$ such that the likelihood \textcolor{black}{Eq.} \eqref{eq:aprrox_stirling} is minimized as 
\begin{align}
\Delta\tau_{\text{opt}} = \mathrm{argmax}_{\Delta\tau}\left(\frac{T}{\Delta\tau}-1\right)\sum_{i\neq j}I_{\Delta\tau}\left(s_i(t+1);s_j(t)\right)\label{eq:opt_tau}
\end{align}

We note that the quantity in the argument of the right hand side of Eq. \eqref{eq:opt_tau} is related to a conventional statistical-testing method called the chi-squared test.
When the dependency is weak, we can write \textcolor{black}{it} as $n^{ij}_{\alpha\beta}/(M-1)=p^{i\cdot}_{\alpha}p^{\cdot j}_{\beta}+\epsilon^{ij}_{\alpha\beta}$ with small values of  $\epsilon^{ij}_{\alpha\beta}$ for each $(\alpha,\beta)\in \{+,-\}^2$.
Then, the gross mutual information $(M-1)I_{\Delta\tau}\left(s_i(t+1);s_j(t)\right)$ is approximated as
\begin{align}
(M-1)I_{\Delta\tau}\left(s_i(t+1);s_j(t)\right)&=(M-1)\sum_{(\alpha,\beta) \in \{+,-\}^2} \frac{n^{ij}_{\alpha\beta}}{M-1}\log\left(1+\frac{\epsilon^{ij}_{\alpha\beta}}{p^{i\cdot}_{\alpha}p^{\cdot j}_{\beta}}\right)\notag\\
&\simeq(M-1)\sum_{(\alpha,\beta) \in \{+,-\}^2} \left(p^{i\cdot}_{\alpha}p^{\cdot j}_{\beta}+\epsilon^{ij}_{\alpha\beta}\right)\left\{\frac{\epsilon^{ij}_{\alpha\beta}}{p^{i\cdot}_{\alpha}p^{\cdot j}_{\beta}}-\frac{1}{2}\left(\frac{\epsilon^{ij}_{\alpha\beta}}{p^{i\cdot}_{\alpha}p^{\cdot j}_{\beta}}\right)^2\right\}\notag\\
&\simeq(M-1)\sum_{(\alpha,\beta)\in \{+,-\}^2}\frac{1}{2}\frac{\left(\epsilon^{ij}_{\alpha\beta}\right)^2}{p^{i\cdot}_{\alpha}p^{\cdot j}_{\beta}}\notag\\
&=\frac{1}{2}\sum_{(\alpha,\beta)\in \{+,-\}^2}\frac{\left(n^{ij}_{\alpha\beta}-(M-1)p^{i\cdot}_{\alpha}p^{\cdot j}_{\beta}\right)^2}{(M-1)p^{i\cdot}_{\alpha}p^{\cdot j}_{\beta}},\label{eq:statistical_quantity_g}
 \end{align}
which holds up to the second order of $\epsilon$.
In the third line we use the constraint $\sum_{(\alpha,\beta) \in \{+,-\}^2}\epsilon^{ij}_{\alpha\beta}=0$ and pick up the terms up to the second order.
The right hand side of Eq.\eqref{eq:statistical_quantity_g} is just the half of Pearson's chi-squared test statistic with $1$ degree of freedom.
This implies that our method can be regarded as a generalization of the chi-square test or g-test \citep{sokalr1981principles}.

\subsection{Screening relevant couplings}

Once the optimal time bin size is decided, \textcolor{black}{applying} the inverse Ising formula to the coarse-grained binary sequence is straight forward.
Calculating the estimated coupling matrix provides us with continuous-valued $J_{ij}^{\mathrm{est}}$ for each pair $(i,j)$ \textcolor{black}{of neurons}.
This continuity \textcolor{black}{may make} results unclear \textcolor{black}{because} it is not easy to distinguish statistically significant couplings from the others.
Therefore, we introduce an additional computational-statistical step \textcolor{black}{\citep{aru2015untangling,xu2017statistical,xu2018inverse}} to extract the relevant couplings from $J^{\mathrm{est}}$.
\textcolor{black}{To prepare,} we generate $L$ randomized time series, each of which is obtained by shuffling  the original coarse-grained sequence  with $\Delta \tau_{\text{opt}}$ in the time direction individually for each element.
For all the series, we calculate the coupling matrices $\{ (J^{\mathrm{est}}_{\mathrm{ran}})_{(r)}\}_{r=1}^{L}$.
Then, for each pair $(i,j)$, we have $L$ reference values $\{ (J^{\mathrm{est}}_{\mathrm{ran}})_{(r)}{}_{ij} \}_{r=1}^{L}$ against the value for the non-randomized data, $J^{\mathrm{est}}_{ij}$.
If \textcolor{black}{the} non-randomized value is relevant, its absolute value is considered \textcolor{black}{as} sufficiently larger than the $L$ reference values.
According to this idea, we accept $J^{\mathrm{est}}_{ij}$ as a relevant coupling only if its absolute value is larger than the $p_{\mathrm{th}}L$ largest value among $\{ |(J^{\mathrm{est}}_{\mathrm{ran}})_{(r)}{}_{ij} |\}_{r=1}^{L}$.
The value $p_{\mathrm{th}}$ is a parameter \textcolor{black}{that controls} the tightness of this criterion and should be small. 
Although we use the simple mean-field formula to obtain $J^{\rm est}$ in this paper, the above procedure can be applied to any other inference algorithms solving the inverse Ising problem. 
Similar data processing methods have been used in other \textcolor{black}{contexts for} spike statistics \citep{bialek2005features,schwartz2006spike}.

This numerical manipulation \textcolor{black}{can induce a great} computational cost when \textcolor{black}{dealing} with a huge amount of data\textcolor{black}{.}
\textcolor{black}{Under such cases it would be} better to make use of the approximation proposed in \citep{terada2018objective}.

\subsection{Symmetric inference}

If the coupling matrix is symmetric, systems are described by an equilibrium distribution in the static manner \citep{kappen1998efficient,tanaka1998mean,sessak2009small,roudi2009ising}.
Hence, we use the mutual information between equal-time states $I_{\Delta\tau}\left(s_i(t);s_j(t)\right)$ instead of $I_{\Delta\tau}\left(s_i(t+1);s_j(t)\right)$ in Eq. \eqref{eq:opt_tau}.
The mean-field inverse formula is replaced as $J^{\text{est}}=A^{-1}-C^{-1}$ \citep{kappen1998efficient,tanaka1998mean}.

To characterize the asymmetric property, we compare the results obtained using symmetric and asymmetric inferences \textcolor{black}{in Sec. \ref{sec:results}}.

\section{Results}\label{sec:results}
\subsection{Synthetic data --- The Izhikevich model}\label{sec:inference_izhikevich}

To confirm the effectiveness of our methods, we \textcolor{black}{first take up the Izhikevich model as an example}.
This is a standard neuronal model \textcolor{black}{that generates} neurobiologically plausible spike sequences \citep{izhikevich2003simple}.
We first set 100 neurons (90 excitatory and 10 inhibitory) on an asymmetric cyclic chain, where each neuron projects synaptic connections \textcolor{black}{and sends signals} to up to \textcolor{black}{three} clockwise neighboring neurons.
The coupling strengths \textcolor{black}{are} drawn from uniform distributions between $5$ and $10$ for the excitatory neurons and between $-20$ and $-10$ for the inhibitory neurons.
The other parameters \textcolor{black}{are} set \textcolor{black}{following the original work} \citep{izhikevich2003simple}.

We also consider the systems on sparse and dense random networks.
For each pair $(i,j)$, except the self pairs, a connection from $j$ to $i$ \textcolor{black}{is} generated with a probability $q$.
\textcolor{black}{In the sparse case, the} coupling strengths \textcolor{black}{are} drawn between $2$ and $3$ for the excitatory neurons and between $-6$ and $-4$ for the inhibitory neurons\textcolor{black}{.}
  \textcolor{black}{In the dense cases, they are} $[0.8/q,0.8/q+1]$ for the excitatory \textcolor{black}{neurons} and $[-2(0.8/q+1),-2(0.8/q)]$ for the inhibitory \textcolor{black}{neurons.}
The other conditions\textcolor{black}{,} except for coupling strengths\textcolor{black}{, are} the same as those in the chain model.

\begin{figure}[h]
 \centering
	\includegraphics[scale=0.23]{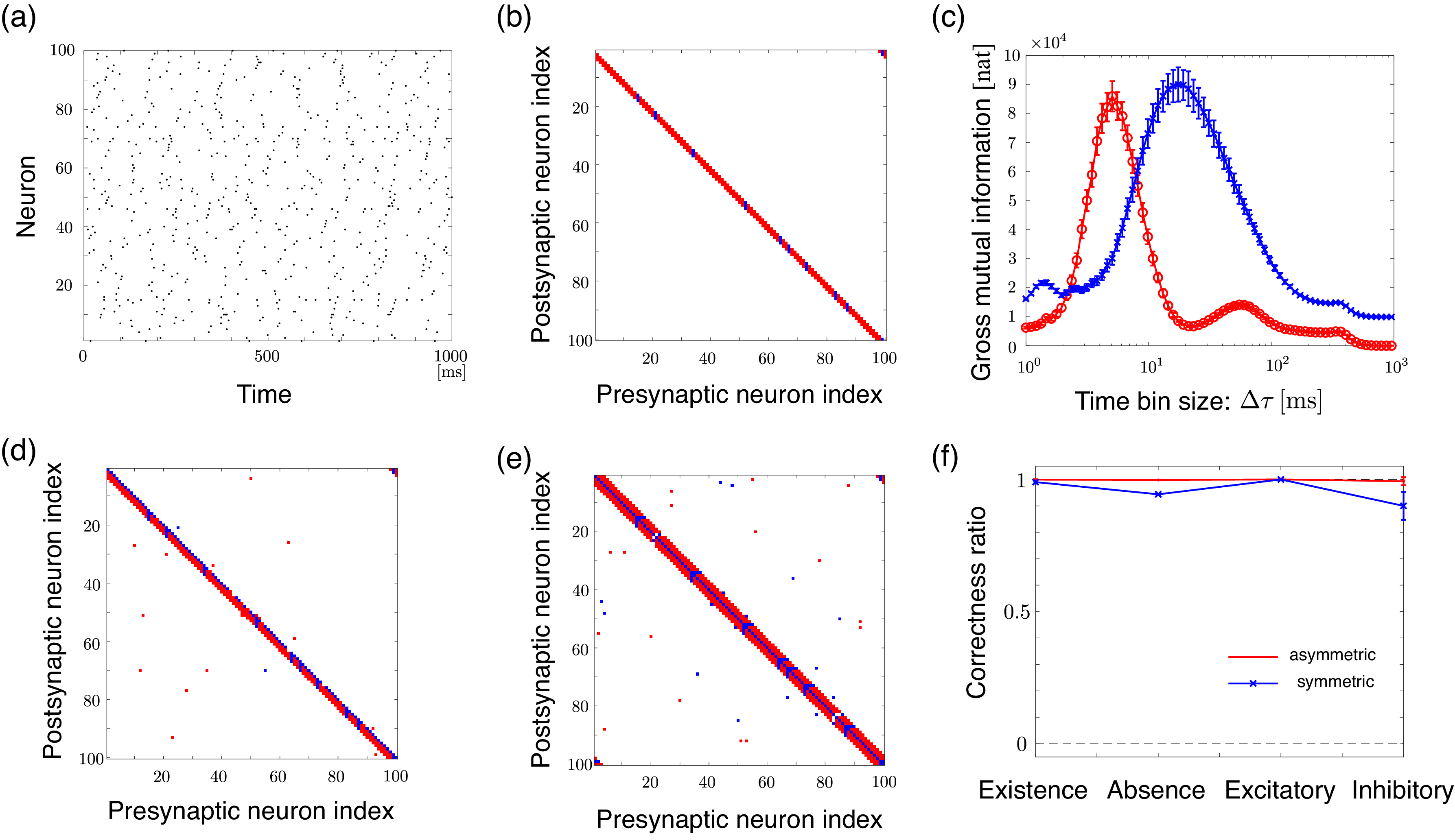}
  \caption{Application to the Izhikevich model on the asymmetric cyclic chain.
  (a) Raster plot during \textcolor{black}{$1\,\mathrm{s}$ with a point plotted for each spike}.
  (b) Network matrix \textcolor{black}{used to generate} the spike data.
  (c) Gross mutual information $(M-1)\sum_{i\neq j}I_{\Delta\tau}\left(s_i(t+1);s_j(t)\right)$ for the asymmetric inference (red) and $(M-1)\sum_{i\neq j}I_{\Delta\tau}\left(s_i(t);s_j(t)\right)$ for the symmetric inference (blue).
  (d) Inferred couplings \textcolor{black}{by the asymmetric inference with $\Delta\tau=5\,\mathrm{ms}$.}
  (e) Inferred couplings \textcolor{black}{by the symmetric inference with $\Delta\tau=15\,\mathrm{ms}$}.
  (f) Conditional ratios of the correct inferences\textcolor{black}{, excudling the self-interactions}.
  \textcolor{black}{In the coupling matrices of (b), (d), and (e), the red and blue elements represent the excitatory and inhibitory couplings, respectively.}
  In (c) and (f) \textcolor{black}{the} error bars represent the \textcolor{black}{standard deviations} over \textcolor{black}{five} different simulations.}
  \label{fig:izhikevich_model}
\end{figure}

The spike trains generated by this model \textcolor{black}{in the cyclic chain} are plotted in Fig. \ref{fig:izhikevich_model} (a) \textcolor{black}{and} the coupling matrix is exhibited in Fig. \ref{fig:izhikevich_model} (b), where the minimal time step is $\Delta t=1\,\mathrm{ms}$ and the time length used for the inference is $T=10^6\,\mathrm{ms}$.
Gross mutual information $(M-1)\sum_{i\neq j}I_{\Delta\tau}\left(s_i(t+1);s_j(t)\right)$ in Eq. \eqref{eq:opt_tau} versus $\Delta \tau$ \textcolor{black}{is shown} by the red curve in Fig. \ref{fig:izhikevich_model} (c), where the error bars show the standard deviations evaluated from \textcolor{black}{five} different simulations.
The red curve produces a nearly unimodal feature with a sharp peak at $\Delta\tau=5\,\mathrm{ms}$, which indicates the unique optimal $\Delta\tau$.
\textcolor{black}{The} nontrivial time scale appears \textcolor{black}{, which may be} due to the difference in the dynamics \textcolor{black}{descriptions} between the Izhikevich and Ising models\textcolor{black}{.}
The Izhikevich model has a process \textcolor{black}{that accumulates} spikes from other neurons to emit a spike\textcolor{black}{, whereas} the Ising has no such process.
\textcolor{black}{Hence, the} appropriate Ising description would need a certain time scale\textcolor{black}{,} presumably longer than the elementary scale \textcolor{black}{for} the Izhikevich model.
This \textcolor{black}{naturally explains the obtained} $\Delta\tau$.
\textcolor{black}{We next create a} binned spike time raster with $\Delta\tau=5\,\mathrm{ms}$, where only $0.23\%$ \textcolor{black}{time} bins contain more than \textcolor{black}{one spike.}
\textcolor{black}{Thereby, we apply} the inverse formula and \textcolor{black}{adopt} only $J_{ij}$ \textcolor{black}{with absolute values exceeding} the threshold or the largest value among the estimates for the $L=1,000$ randomized time series\textcolor{black}{.}
\textcolor{black}{This} corresponds to the criterion $p_{\text{th}}=0.001$ \textcolor{black}{and yields} the relevant couplings shown in Fig. \ref{fig:izhikevich_model} (d).
The original asymmetric structure with the excitatory and inhibitory couplings \textcolor{black}{is sufficiently recovered, despite that} the models used for generation and inference are different.
The \textcolor{black}{self-interactions} tend to be inferred as inhibitory couplings\textcolor{black}{, although they are absent} in the original system\textcolor{black}{.
This} may describe a refractory property of the neurons.
As a comparison, we apply the symmetric inference procedure to the same data,  the results of which are shown \textcolor{black}{by} the blue curve in Fig. \ref{fig:izhikevich_model} (c), \textcolor{black}{which exhibits} a larger optimal $\Delta\tau$\textcolor{black}{, and by the matrix in Fig. \ref{fig:izhikevich_model} (e)}.
To express the regularity of the asymmetric model using the symmetric model, it is necessary to merge successive bins \textcolor{black}{to describe} events in neighboring time steps as simultaneous events.
This may be why the optimal bin size for the symmetric model is considerably larger than for the asymmetric model.
The estimated couplings with $\Delta\tau=15\,\mathrm{ms}$ in Fig. \ref{fig:izhikevich_model} (f) are localized around the diagonal line\textcolor{black}{,} but distributed \textcolor{black}{on} both sides, while the true couplings used for the simulation are \textcolor{black}{unidirectional}.
We also show the conditional ratios \textcolor{black}{for} correctness under conditions of existence, absence, excitatory and inhibitory couplings in Fig. \ref{fig:izhikevich_model} (f).
Both models \textcolor{black}{successfully identify} the connections with their signs, although the symmetric inference \textcolor{black}{model} has larger false positive rates.

\begin{figure}
  \centering
	\includegraphics[scale=0.23]{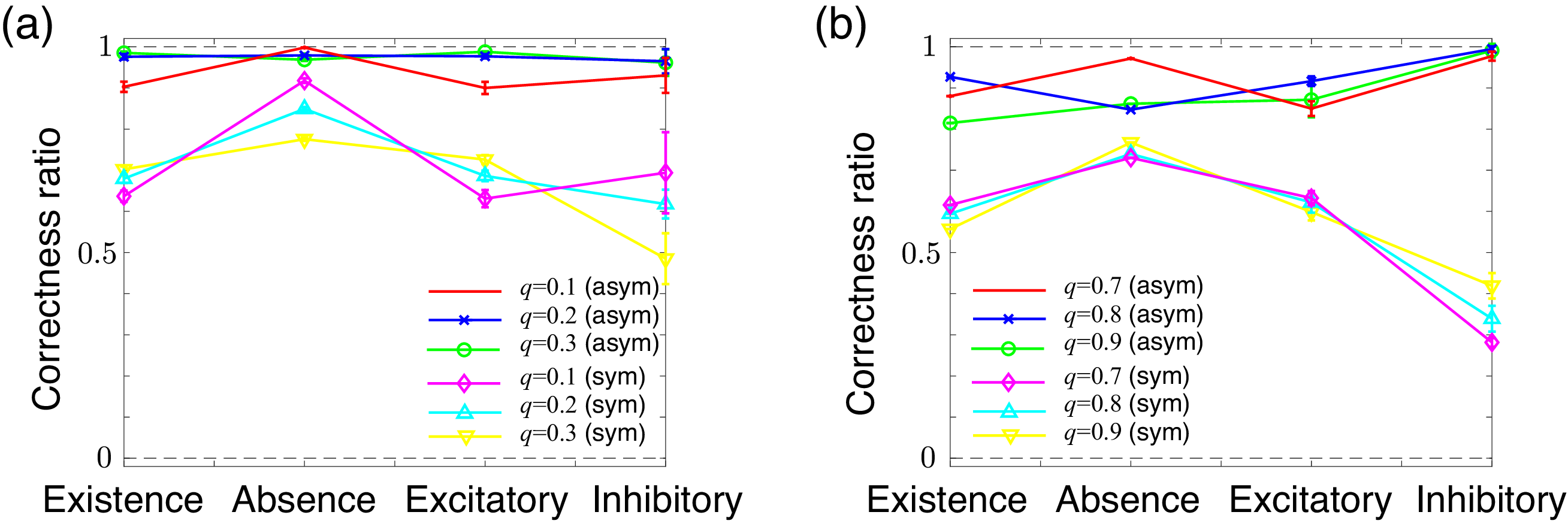}
  \caption{Application to the Izhikevich model on the asymmetric random network.
  (a) \textcolor{black}{Sparse} and (b) dense connectivities, where $q$ denotes the connection probability.
  Conditional ratios of the correctness\textcolor{black}{, excluding the self-interactions,} in the asymmetric and symmetric inferences.
  The \textcolor{black}{error} bars represent the \textcolor{black}{standard deviations} over \textcolor{black}{five} simulations.}
  \label{fig:izhikevich_network}
\end{figure}

To scrutinize the conditions where our inference procedure \textcolor{black}{works} well, \textcolor{black}{the inference methods are studied using sparse and dense random networks}.
The obtained optimal bin sizes \textcolor{black}{are} not so different from the ones \textcolor{black}{in} the chain model, \textcolor{black}{so we adopt} the same values.
The conditional ratios of the correctness in the asymmetric and symmetric inferences are shown in Fig. \ref{fig:izhikevich_network}.
The asymmetric model has much higher expressive power and\textcolor{black}{, hence,} higher performance.
These results highlight the broad applicability of the proposed inference procedure.

\subsection{Experimental data --- \textit{in vitro} neuronal networks}

\begin{figure}
  \centering
	\includegraphics[scale=0.3]{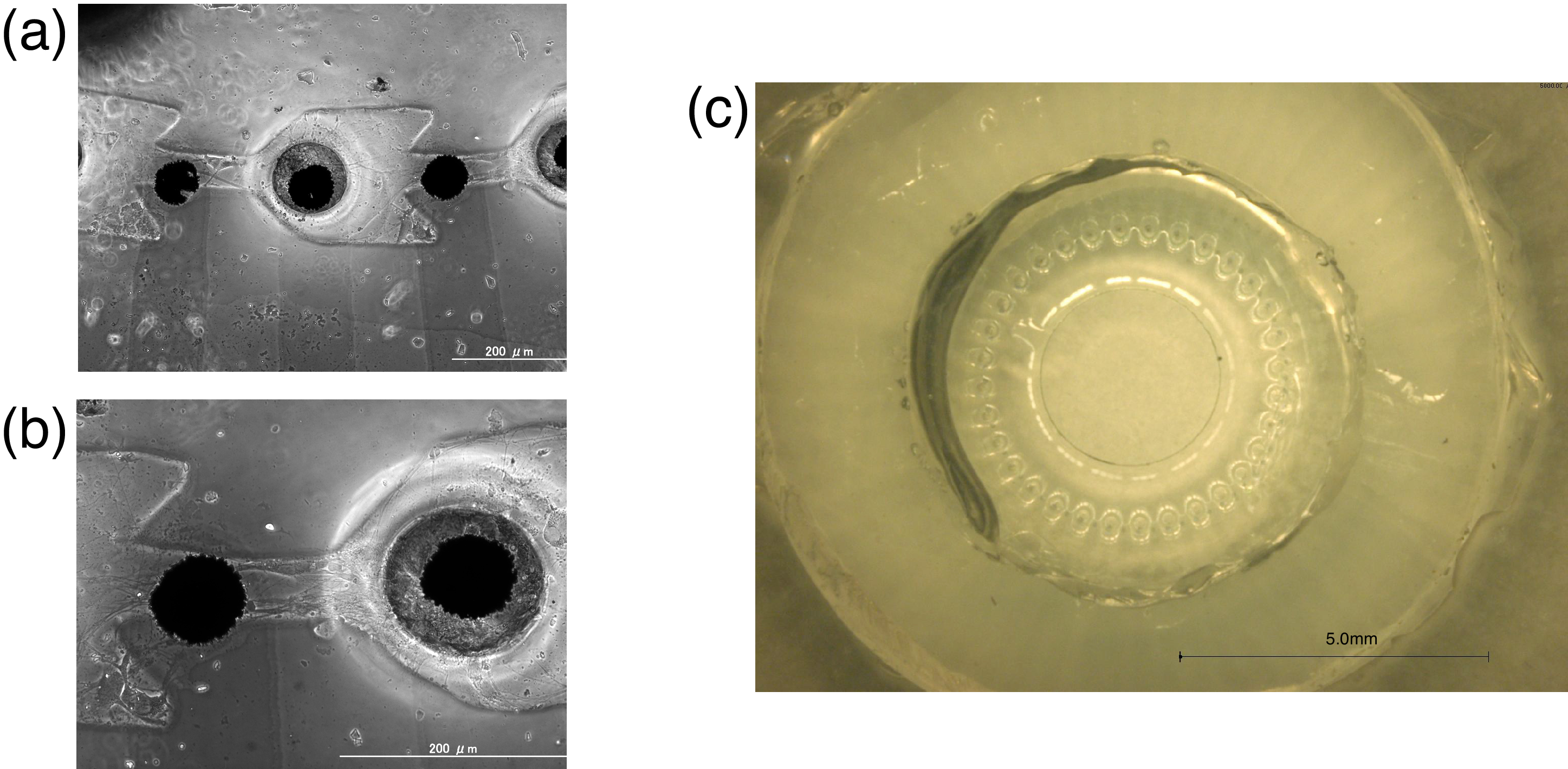}
  \caption{Pictures of the cultured system.
  (a,b) Neurons were cultured in micro-chambers designed to guide the axons in one-direction.
  The black dots denote micro electrodes.
  (c) The entire experimental equipment setup \textcolor{black}{constructed so that the neurons and their synapses form a physical asymmetric chain network}.}
  \label{fig:experimental_condition}
\end{figure}

We study a cultured neuronal system introduced in \citep{isomura2015signal} to demonstrate the applicability of our methods \textcolor{black}{in} real systems.
Rat cortical neurons were cultured in a \textcolor{black}{microwell plate} so that they were likely to \textcolor{black}{asymmetrically connect} to clockwise neighboring cells\textcolor{black}{, as shown} in Fig. \ref{fig:experimental_condition}\textcolor{black}{.}
\textcolor{black}{This} provides a similar condition to that of the Izhikevich model on the chain\textcolor{black}{, which is studied in Sec. \ref{sec:inference_izhikevich}.}
Spontaneous spiking \textcolor{black}{neuron activity was} recorded from 64 electrodes of a multi-electrode array\textcolor{black}{.}
\textcolor{black}{The} measurement time \textcolor{black}{was} $\Delta t=40\,\mathrm{\mu s}$ and the time length \textcolor{black}{of the data} used here was $120\,\mathrm{s}$.
Spike sorting \citep{takekawa2010accurate} was subsequently applied to the \textcolor{black}{recorded} data and 100 neurons were identified.

Generally, we do not precisely know the underlying connectivity in real neuronal systems, which provides uncertainty regarding the accuracy of inference methods. 
However, this favorable situation allows us to easily confirm the validation of the inference methods due to the physical circular restriction of the synaptic connections.

\begin{figure}
  \centering
	\includegraphics[scale=0.23]{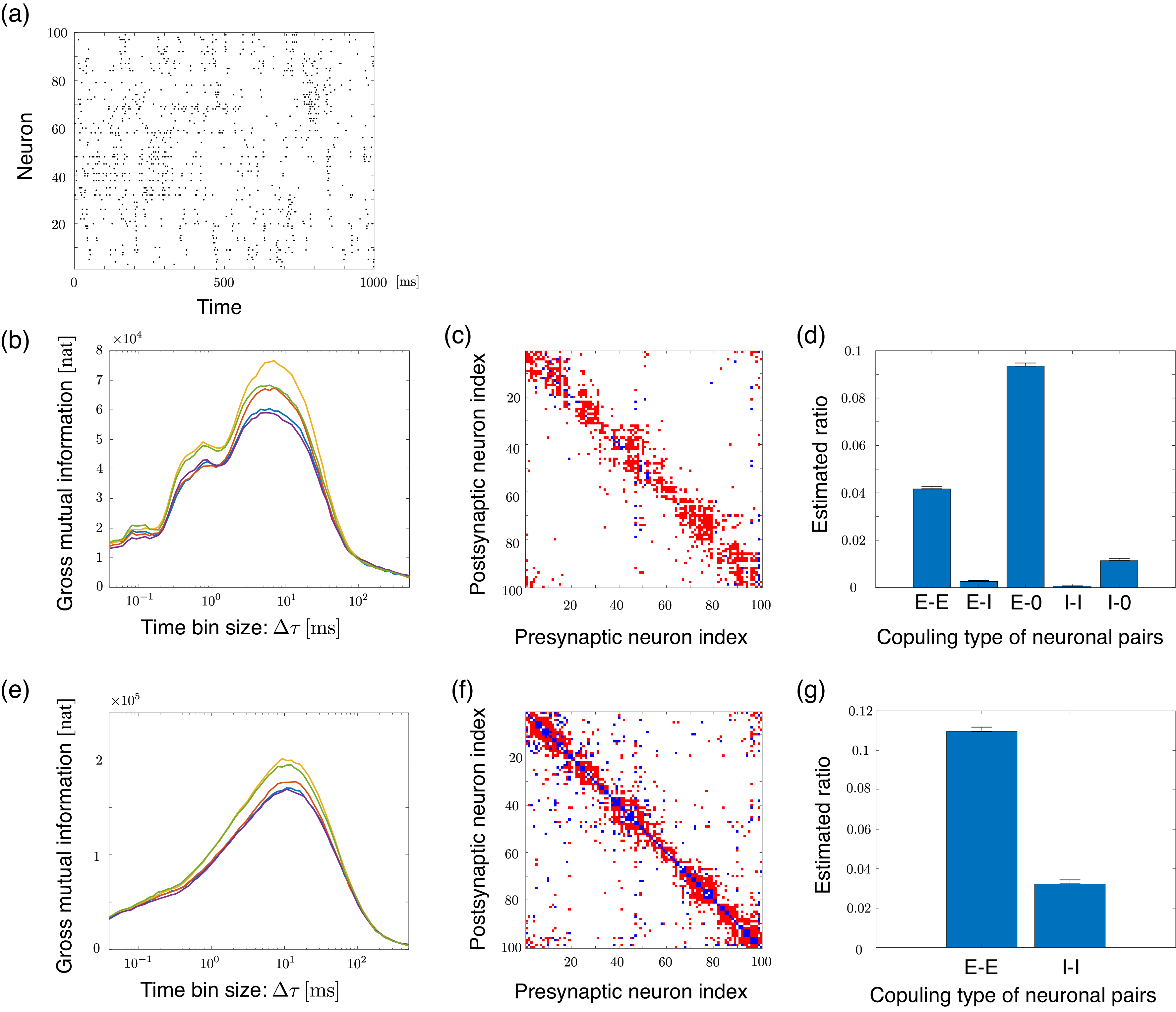}
  \caption{Application to a cultured-neuronal system.
  (a) Raster plot of the spontaneous activity during $1\,\mathrm{s}$.
  (b) Gross mutual information in Eq. \eqref{eq:opt_tau}, where the results for \textcolor{black}{five} different \textcolor{black}{$1\,\mathrm{s}$ intervals} are shown. 
  (c) Inferred couplings by the asymmetric inference with $\Delta\tau=4 \,\mathrm{ms}$\textcolor{black}{.}
  (d) Ratios of estimated coupling types,excluding the absence type.
  The error bars indicate the standard deviation over the five periods.
  (e-g) Corresponding panels for the symmetric inference method using the same data.
  In (f) the network is inferred with a time bin size of $\Delta\tau=10 \,\mathrm{ms}$.
  In the estimated coupling matrices of (c) and (f), the red and blue elements represent the excitatory and inhibitory couplings, respectively.}
  \label{fig:clutured}
\end{figure}

The spike raster plot during \textcolor{black}{$1\,\mathrm{s}$} is exhibited in Fig. \ref{fig:clutured} (a).
Gross mutual information for \textcolor{black}{five} different periods is plotted in Fig. \ref{fig:clutured} (b).
An almost unimodal shape is also observed, where the locations of the peaks are robust.
We set $\Delta\tau=4\,\mathrm{ms}$ (where $1.81\%$ spike bins contain more than one spike), which almost maximizes the mutual information, and proceed to perform the inverse asymmetric Ising inference and pick up the relevant interactions, similar to the Izhikevich models.
\textcolor{black}{The} bold diagonal outline appears \textcolor{black}{Fig. \ref{fig:clutured} (c)}\textcolor{black}{, where the neuronal indices are set following the locations of the corresponding electrodes as in \citep{isomura2015signal}.}
This indicates a chain-like structure, which precisely reproduces the experimentally designed neuronal network structure \citep{isomura2015signal}.
Therefore, our statistical inference captures the underlying physiological couplings.
The ratios of the estimated coupling-types between pairs\textcolor{black}{, except for the absent pairs,} are shown in \textcolor{black}{Fig. \ref{fig:clutured}(d)}.
In this setting, the majority of the estimated connections are the one-way paths with excitatory couplings while some excitatory-excitatory couplings and inhibitory one-way paths are also estimated.
These results \textcolor{black}{reproduce the actual asymmetric structure and are consistent with the physiological property of \textit{in vitro} networks, such as the ratio of the excitatory and inhibitory neurons that has been previously reported \citep{marom2002development}}. 

For comparison, we also plot the counterpart results obtained from the conventional symmetric inference methods in Fig. \ref{fig:clutured} (e-g). 
A similar chain-like network structure is generated; 
however, the coupling plot Fig. \ref{fig:clutured} (f) tends to be noisier and their finer coupling structures are missed. 
These results highlight the effectiveness of our methods with the use of the asymmetric model. 

\section{Discussion}\label{sec:discussion}
\subsection{Robustness of bin-size determination: comparison with cross correlation}\label{sec:crosscorrelation}

Our statistical study confirmed that the proposed determination of the optimal bin size based on Eq. \eqref{eq:opt_tau} is adequately robust, as shown in Fig. \ref{fig:izhikevich_model} (c) and Fig. \ref{fig:clutured} (b, e).
Here, we examine cross-correlation as a comparison, which is another quantity commonly used for determining the characteristic time scale of the data \citep{ostojic2009how}.

\begin{figure}
  \centering
  \hspace*{-1.5cm}
	\includegraphics[scale=0.23]{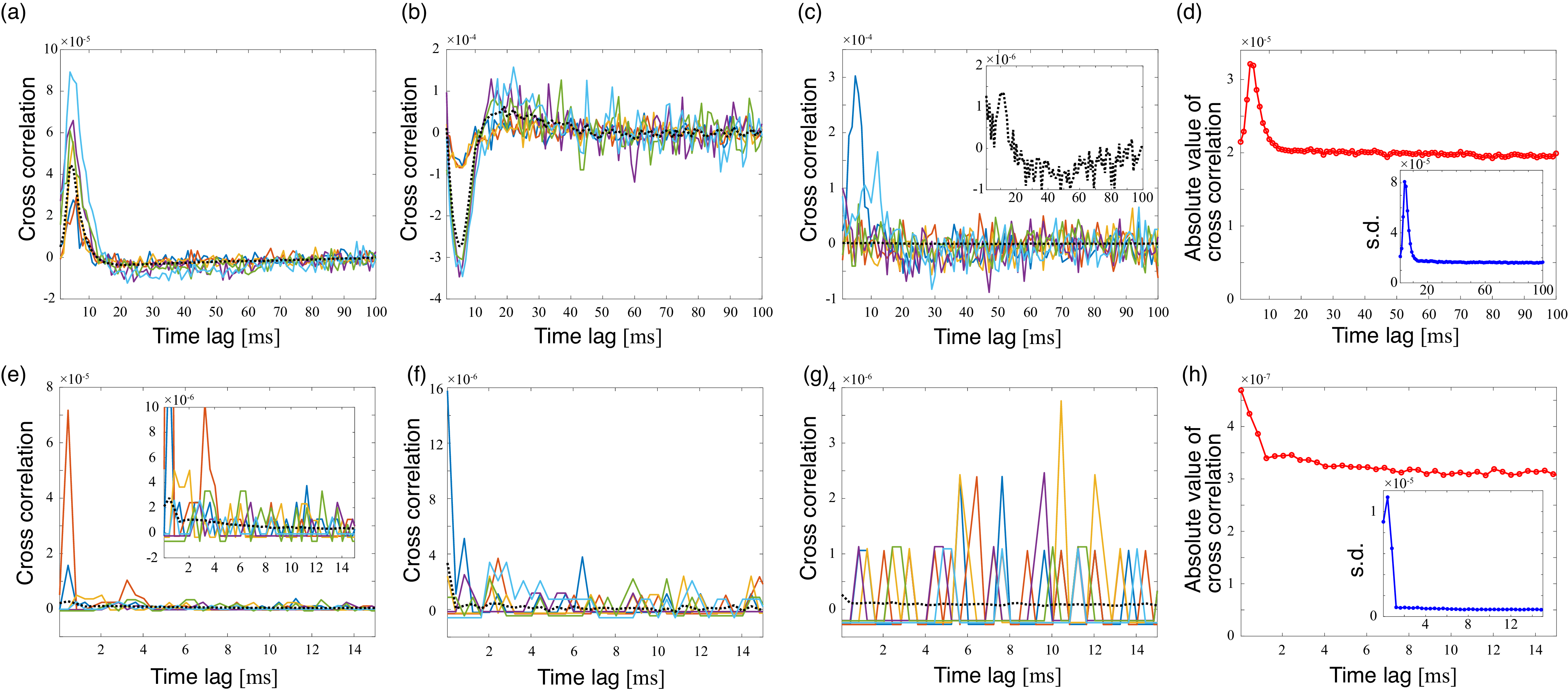}
  \caption{\textcolor{black}{Cross correlations between neuronal pairs using data from (a, b, c) the Izhikevich model and (e, f, g) the \textit{in vitro} neuronal system and (d,h) the means of the absolute values of the cross correlations.
  Cross correlations for six randomly chosen pairs with (a) excitatory, (b) inhibitory, and (c) no couplings, where the black dotted lines represent the means over all the excitatory (270 pairs), inhibitory (30 pairs), and absent couplings, respectively.
  The inset in (c) represents  the mean value on the different scale. 
  (d) Mean of the absolute value of the cross correlations over all pairs in the Izhikevich model, where the inset represents the standard deviations over all pairs.
  Correlations of six randomly chosen pairs with estimated (e) excitatory, (f) inhibitory, and (g) no couplings, where the black dotted lines represent the means as in the Izhikevich case and the estimated excitatory and inhibitory pairs are 862 and 62, respectively. 
  The inset in (e) represents  the cross correlation and the mean value on the different scale. 
  (h) Absolute mean of all pairs in the cultured system, where the inset represents the standard deviations over all pairs.}
}
  \label{fig:cross_correlation}
\end{figure}

The cross-correlations in the Izhikevich model and the cultured system are shown in Fig. \ref{fig:cross_correlation}, \textcolor{black}{which uses} the same data as in Figs. \ref{fig:izhikevich_model} and \ref{fig:clutured}, respectively.
We show the correlations in the  excitatory (Fig. \ref{fig:cross_correlation} (a)), inhibitory (Fig. \ref{fig:cross_correlation} (b)), and no coupling (Fig. \ref{fig:cross_correlation} (c)) pairs, and the mean of their absolute values in the Izhikevich model (Fig. \ref{fig:cross_correlation} (d)).
In the Izhikevich model, the characteristic time scale at approximately $5\,\mathrm{ms}$ appears in the excitatory and inhibitory couplings, but is not clearly observed in the case of no coupling cases.
We state that this time scale is extremely close to that obtained by Eq. \eqref{eq:opt_tau} in the Results section, which implies that both of our method and cross-correlation method can provide accurate inference in this case.
Conversely, Fig. \ref{fig:cross_correlation} (e-h) shows the cultured system and demonstrates no characteristic time scale, except the time resolution.
However, as shown in Fig. \ref{fig:clutured} (b), Eq. \eqref{eq:opt_tau} can effectively detect the characteristic time scale.
Thus, we conclude that the applicability of our methods is broader than the cross-correlation-based method.

\begin{figure}[h]
  \centering
	\includegraphics[scale=0.25]{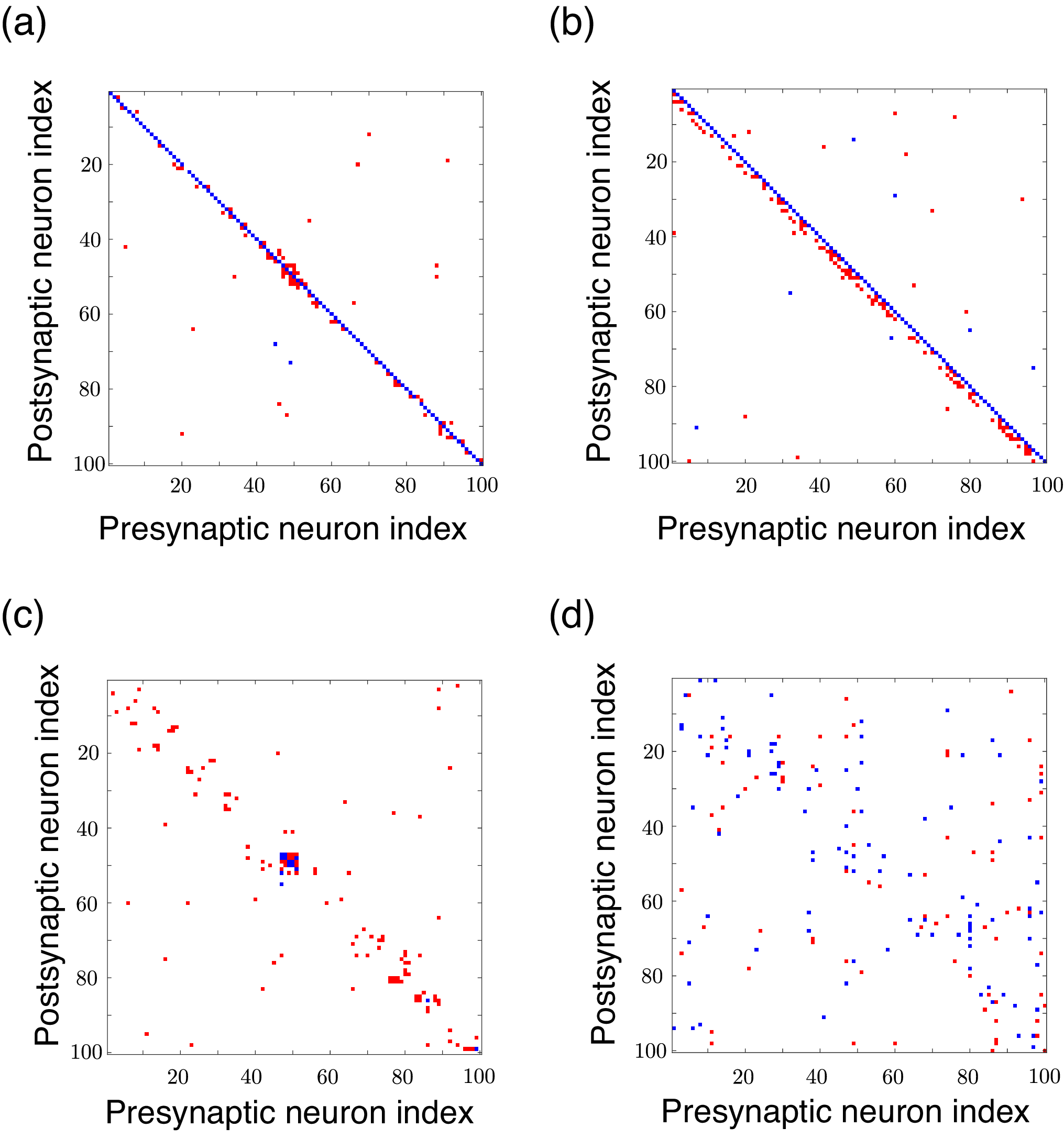}
  \caption{Inferred coupling matrices with \textcolor{black}{non-optimal time bin sizes}.
  The Izhikevich model using (a) $1\,\mathrm{ms}$ and (b) $20\,\mathrm{ms}$.
  The cultured system using (c) $0.04\,\mathrm{ms}$ (time resolution) and (d) $100\,\mathrm{ms}$.
  The other conditions are the same as those in Fig. \ref{fig:izhikevich_model} (d) and Fig. \ref{fig:clutured} (c).
}
  \label{fig:nonoptimal_bins}
\end{figure}

Moreover, we demonstrate the  inference results using non-optimal bin sizes in the Izhikevich and cultured systems in Fig. \ref{fig:nonoptimal_bins}.
Fig. \ref{fig:nonoptimal_bins} (a, b) corresponds to the results in Fig. \ref{fig:izhikevich_model} (d) for the Izhikevich model and Fig. \ref{fig:nonoptimal_bins} (c, d) corresponds to Fig. \ref{fig:clutured} (c) for the cultured system, where discretizations with shorter and longer non-optimal sizes are used.
This result implies that using optimal bin sizes results in clearer inference than using the smaller and larger non-optimal bin sizes.

\subsection{Finite-size bias of mutual information}

We used the gross mutual information of binarized time series to determine the optimal time bin size for coupling inference in Section \ref{sec:results}.
In the large limit of the data size, we can obtain accurate estimate for the gross mutual information from the observed data.
However, in real situations the data size is finite, and hence it is important to assess the finite-size bias for that quantity.
Thus, we calculate the bias of the mutual information $I(s_i(t+1);s_j(t))$ induced by the finiteness of the used data size by using a resampling method.

\begin{figure}
  \centering
	\includegraphics[scale=0.25]{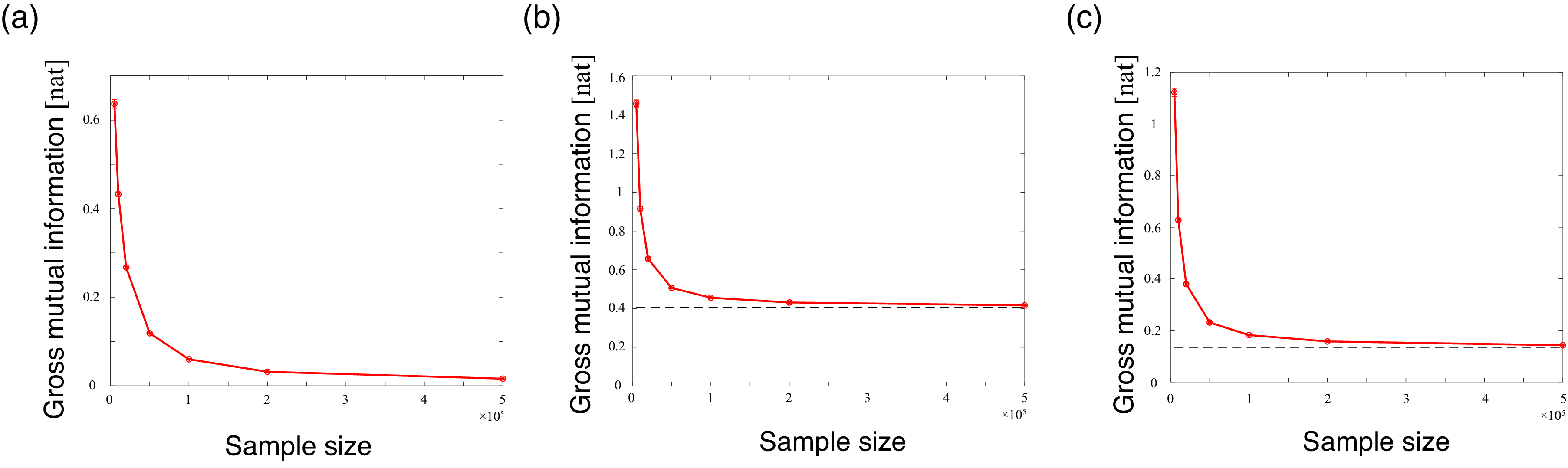}
  \caption{\textcolor{black}{Finite size bias of the gross mutual information estimated by the resampling method. 
  The values of $\{p_{++},p_{+-},p_{-+},p_{--}\}$ obtained from the Izhikevich model simulations at (a) $\Delta \tau=1{\rm ms}$, (b) $5{\rm ms}$, and (c) $20{\rm ms}$ are used. 
  The points and error bars denote the average and standard deviation, respectively, over 20 independent simulations. 
  The dashed lines represent the gross mutual information in the infinite sample limit.}}
  \label{fig:mi_bias}
\end{figure}

\textcolor{black}{
Given the binarized data $\left\{s_i(t)\right\}_{i=1,\cdots,N;t=1,\cdots,T}$, we have the empirical one-step joint probability $\{p_{++}^{ij},p_{+-}^{ij},p_{-+}^{ij},p_{--}^{ij}\}$, which also generates the marginalized probabilities $p_{+}^{i\cdot},p_{-}^{i\cdot},p_{+}^{\cdot j},p_{-}^{\cdot j}$ for all $(i,j)$ pairs.
Using these probabilities, we generate a binary sequence of finite sizes, where we randomly choose a previous state from the original binarized data.
We calculate the gross mutual information for these resampled sequences, which describes the finite-size bias for the data.
The results are shown by the red curves in Fig. \ref{fig:mi_bias}.
Here we use the values of $\{p_{++},p_{+-},p_{-+},p_{--}\}$ obtained from the Izhikevich model simulations in Fig. \ref{fig:izhikevich_model} at $\Delta \tau=1{\rm ms}$, $5{\rm ms}$, and $20{\rm ms}$.
Although the gross mutual information tend to be positively biased at finite sample sizes, we observe that they smoothly approach the dashed line corresponding to the infinite size limit. 
This figure also suggests that the deviation due to the sample finiteness from the infinite size limit is small enough in our present setting.
This is exemplified by Fig. \ref{fig:mi_bias} (c), where the sample size is $M=5\times 10^4$, the lowest among the examined $\Delta \tau$ range. 
Therefore, our gross mutual information estimates and the optimal time bin sizes obtained in Section \ref{sec:results} are considered to be accurate.} 

\subsection{Comparison to other regularization method}\label{sec:regularization}

This paper proposes a method for screening significant couplings from the estimates.
Several regularization methods to avoid overfitting have been proposed in other works \citep{zeng2014l1,decelle2015inference,bulso2016sparse}.
Among them, some sparsity-inducing regularizations also work as an effective screening method, which are of interest to compare with our method. 
Therefore, we examine the $\ell_1$ regularization as a representative example of sparsity-inducing regularizations.

\begin{figure}
  \centering
	\includegraphics[scale=0.3]{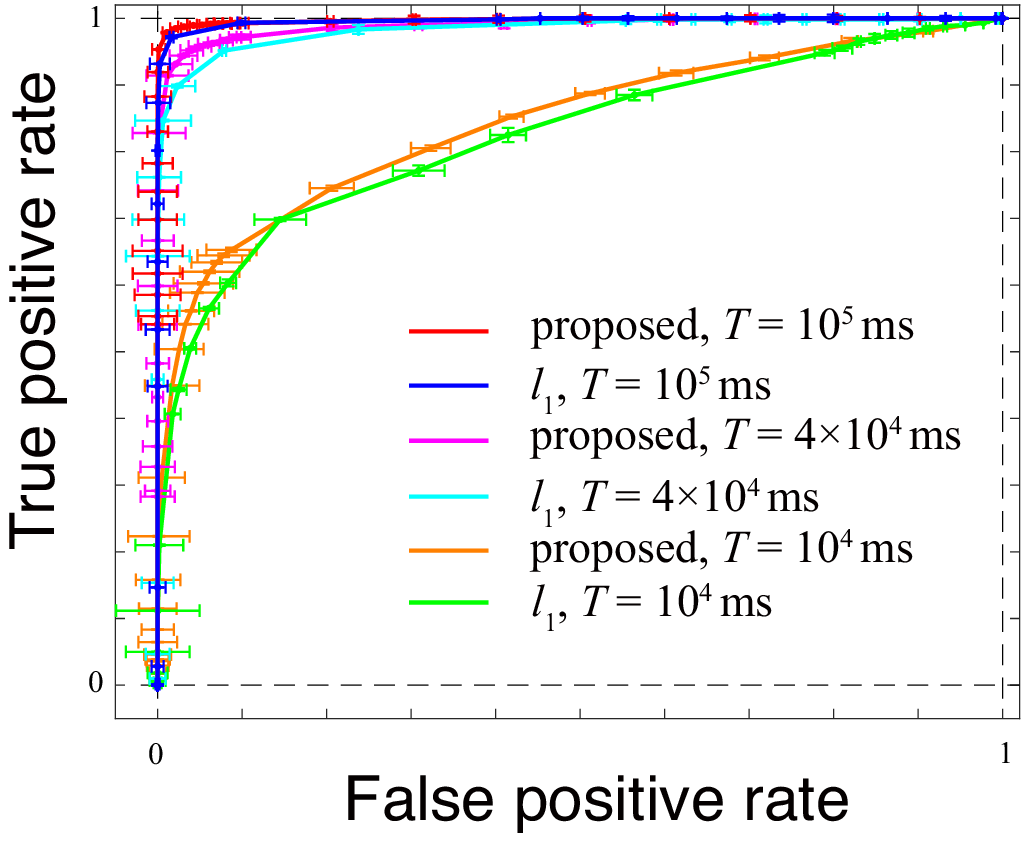}
  \caption{\textcolor{black}{ROC curves  by the proposed and $\ell_1$ regularization methods for different data sizes in the Izhikevich model.}
  \textcolor{black}{The error bars represent the standard deviations over five simulations.}}
  \label{fig:l1}
\end{figure}

For the numerical experiment using the $\ell_1$ regularization, we employ the \textit{lassoglm} package in MATLAB, which estimates parameters based on the maximum likelihood in the generalized linear model. 
We obtain the ROC curves by changing the level of significance values for the proposed method and the coefficient of the regularizer for the $l_1$ regularization, which are shown for several data sizes in Fig. \ref{fig:l1}.
For comparison, we plot the ROC curve based on our method, where we use the semi-analytical method to compute the values at very small FP and TP \citep{terada2018objective} to avoid the huge computational cost induced by the randomization operation.
This result shows that our method reconstructs ground-truth couplings as accurate as the $\ell_1$ regularization does.
The advantage of our method is that it requires a lower computational cost compared to other methods and does not require any specific prior-assumptions.

\subsection{Goodness of fit of the kinetic Ising model}\label{sec:goodness}

We investigate the goodness of fit achieved by the kinetic Ising model.
Although we used the mean-field (MF) approximation as the inference algorithm in Section \ref{sec:results}, we here use the maximum likelihood (ML) method to evaluate the expressive ability of the kinetic Ising model (the generalized linear model) itself and to assess the MF approximation accuracy.
To perform the ML method, we use the {\it glmfit} package in MATLAB.
The \textit{glmfit} package considers the spike train data as $\sigma_i\in\left\{0,1\right\}$, where 1 indicates the existence of a spike and 0 indicates no spike, during direct application of the package code.
We then rephrase the result in the Ising description as $s_i=\pm1$ through a simple transformation of the variables.
To evaluate the statistics in the kinetic Ising model, we generate the binary data using the transition probability which is conditioned by the original data at the previous time step.

\begin{figure}
  \centering
	\includegraphics[scale=0.25]{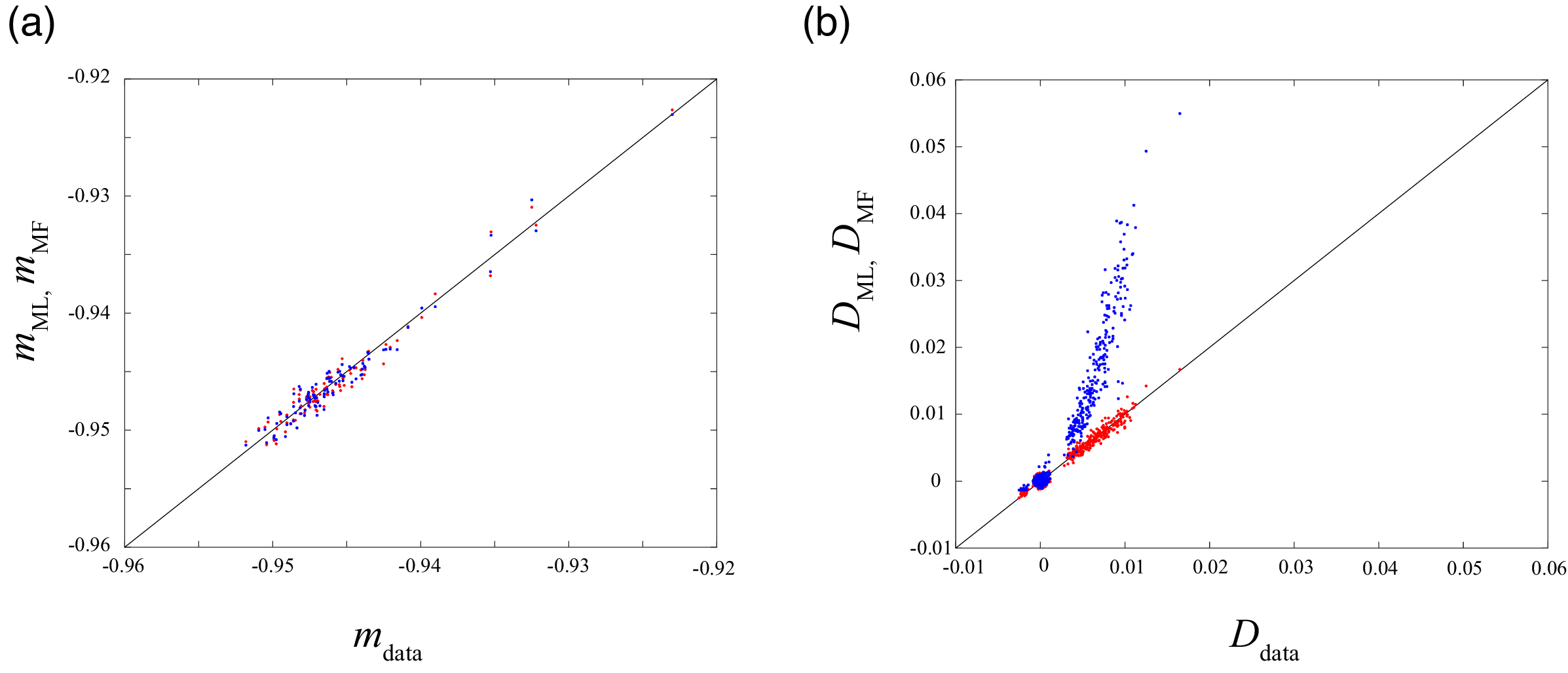}
  \caption{Statistical reconstruction of the synthetic data using the kinetic Ising model.
  \textcolor{black}{(a) Means and (b) different time correlations obtained from the original data in the Izhikevich model \textcolor{black}{compared to the data generated} by the kinetic Ising model \textcolor{black}{using parameters} obtained from the maximum likelihood of the red cross points and \textcolor{black}{the} mean-field inverse formula of the blue x's.}}
  \label{fig:statistics}
\end{figure}
To study the goodness of fit of the model\textcolor{black}{,} we conduct cross validations \textcolor{black}{for} the synthetic and experimental data.
We separate their spike trains into two parts \textcolor{black}{with the same lengths:}
one \textcolor{black}{is} used to infer the couplings \textcolor{black}{and} the other \textcolor{black}{is} used to evaluate the statistics \textcolor{black}{generated by} the model.
Using the coarse-grained data with the optimal time \textcolor{black}{bins} in the Izhikevich model, we calculate the mean and different-time correlation and compare them with those of the data generated by the kinetic Ising model\textcolor{black}{.} 
\textcolor{black}{In Fig. \ref{fig:statistics} (a) and (b) the red crosses} show the means and correlations of the original data from the Izhikevich model and the generated data by the \textcolor{black}{kinetic Ising} model with the ML parameters.
Good consistency is observed between the two statistics, which confirms that the \textcolor{black}{kinetic Ising model} can almost completely \textcolor{black}{recover} the \textcolor{black}{data's} one-step transition statistics \textcolor{black}{when the ML method is used.
The counterparts of the MF approximation in the kinetic Ising model are shown by the blue x points.}
Although the MF inference loses finer statistical \textcolor{black}{properties compared to the ML method, especially regarding the correlation,} the sign property is retained\textcolor{black}{.}
\textcolor{black}{We interpret} that this is a source for the accurate discrimination of the excitatory, inhibitory, and absent couplings demonstrated in the Results section.

\begin{figure}
  \centering
	\includegraphics[scale=0.25]{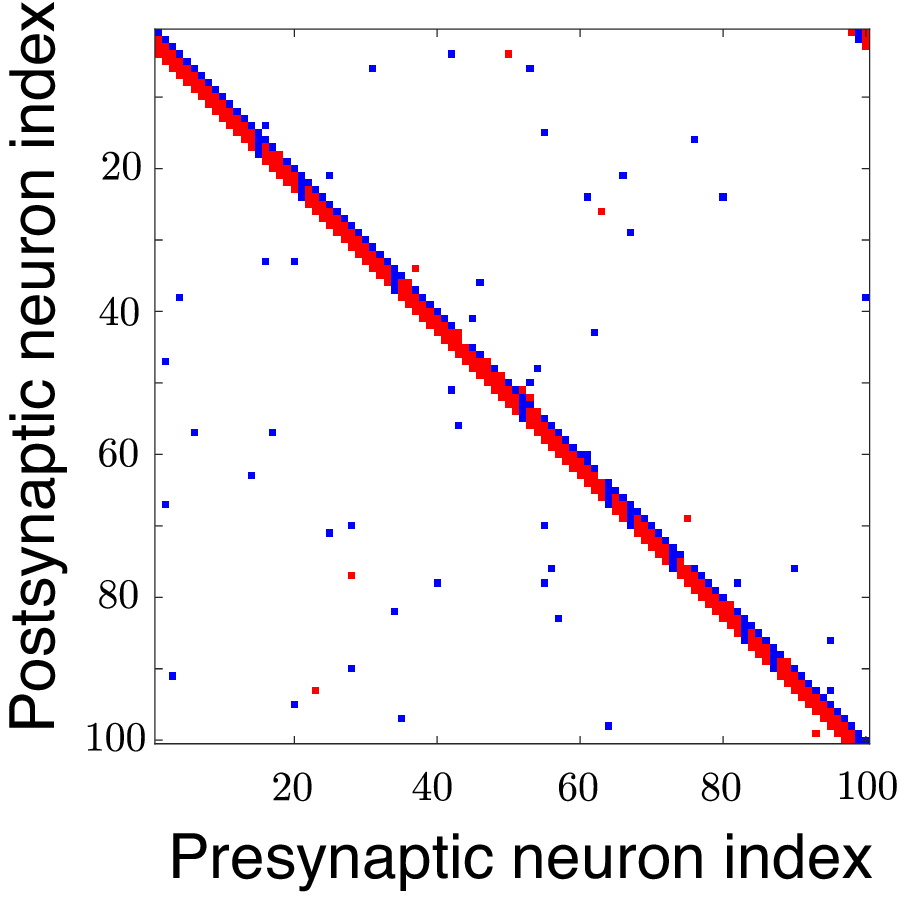}
  \caption{Inferred couplings \textcolor{black}{generated} by the maximum likelihood method. 
 The conditions are the same \textcolor{black}{as those in Fig. \ref{fig:izhikevich_model} (d), except for the algorithm}.}
  \label{fig:couplings}
\end{figure}
\begin{table}[h]\label{fig:performance}
\scalebox{0.75}[0.7]{
  \begin{tabular}{|c||c|c|c|c||c|} \hline
    Methods & Correct Ratio: Existence & Absence & Excitatory & Inhibitory & Time [s] \\ \hline
    Mean Field & $0.9993\pm0.0015\,(\mathrm{s.d.})$ & $0.9979\pm0.0006$ & $1\pm0$ & $0.9933\pm0.0149$ & $55344$ \\
    Maximum Likelihood & 1 & 0.9944 & 1 & 1 & $529772$ \\ \hline
  \end{tabular}}\\
    \caption{Accuracy and computational cost of the inference of the mean-field approximation and direct maximization of the likelihood in the Izhikevich model.}
\end{table}
Although the direct ML inference requires \textcolor{black}{a much greater} computational cost than that with the MF method, we \textcolor{black}{are able} to apply our screening method to the ML estimator. 
The result is shown in Fig. \ref{fig:couplings}.
We observe no significant difference between this and \textcolor{black}{Fig. \ref{fig:izhikevich_model} (d)}.
Moreover, we quantitatively demonstrate the accuracy and computational cost of the ML and MF inferences  in Table 1.
The computational cost is addressed for screening process with $1,000$ surrogate data manipulation.
We conclude that MF inference drastically saves the computational cost \textcolor{black}{and} maintains the inference accuracy compared to the ML inference.

\begin{figure}[h]
  \centering
	\includegraphics[scale=0.25]{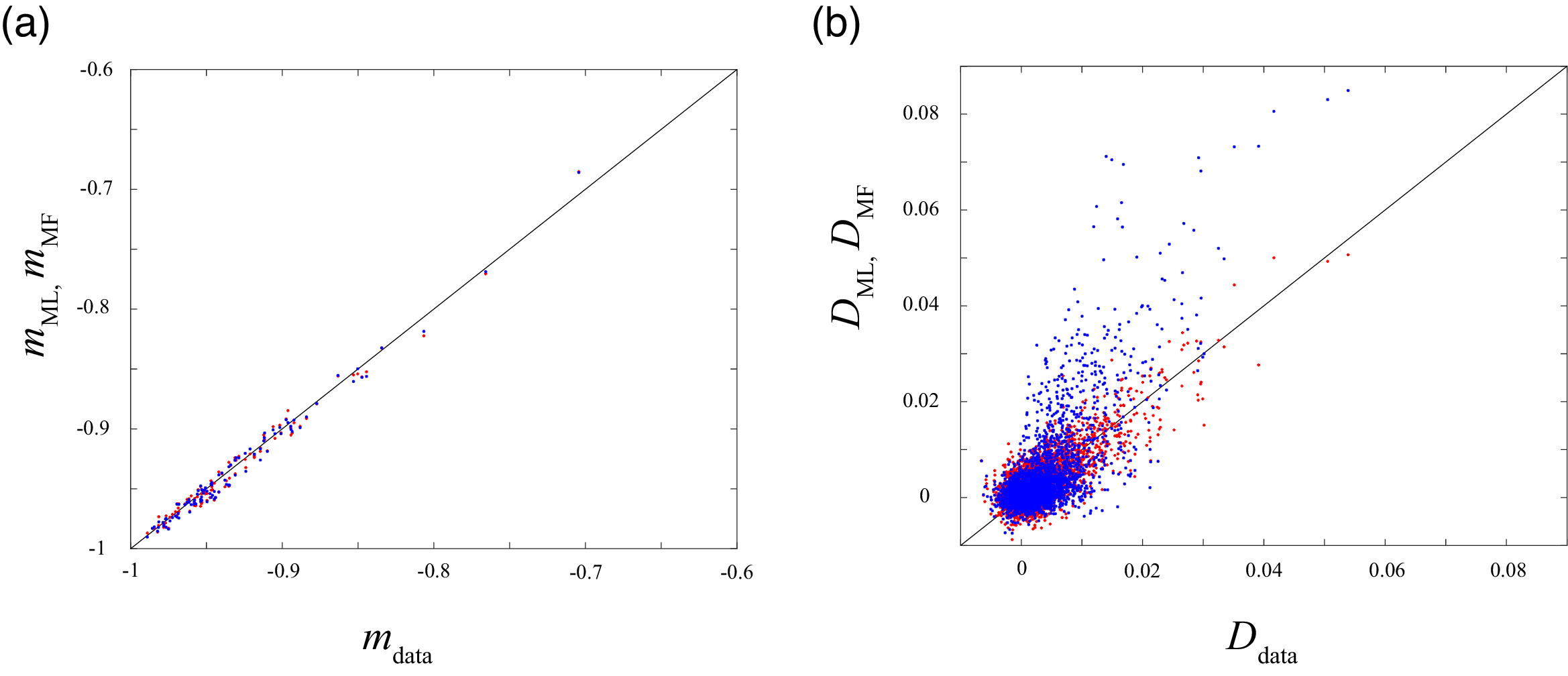}
  \caption{\textcolor{black}{Statistical reconstruction of the experimental data using the kinetic Ising model}.
  \textcolor{black}{(a) Means and (b) different time correlations from the original \textcolor{black}{experimental} data \textcolor{black}{compare to the data generated} by the kinetic Ising model \textcolor{black}{using parameters} obtained from the maximum likelihood of the red cross points and \textcolor{black}{the} mean-field inverse formula of the blue x's.}}
  \label{fig:statistics_spontaneous}
\end{figure}

Additionally, we evaluate the goodness of fit in the experimental data shown in Fig. \ref{fig:statistics_spontaneous}.
Although the fine statistics do not recover, especially \textcolor{black}{in the MF correlation}, most of them retain the \textcolor{black}{tendencies} of the sign patterns\textcolor{black}{, similar to what is} observed in the Izhikevich model.

\subsection{Fundamental limitation by dominant long-range collective modes}\label{sec:limitation}

Finally, we state that the presence of dominant long-range modes in systems can hide local statistics and degrade the inference accuracy.
As recently reported in \citep{das2019systematic}, most inference methods fail to reconstruct neuronal couplings in the presence of long-range modes.
Indeed, if such modes hide in data and contaminate the statistics directly related to couplings, it is \textcolor{black}{very} difficult to infer the connectivity.
However, \textcolor{black}{using} statistical quantities, we can assess the presence or absence of these long-range modes\textcolor{black}{, allowing us to avoid these situations in advance}.
We use a recurrent network model introduced by \citep{das2019systematic} to demonstrate this.

\begin{figure}[h]
  \centering
	\includegraphics[scale=0.19]{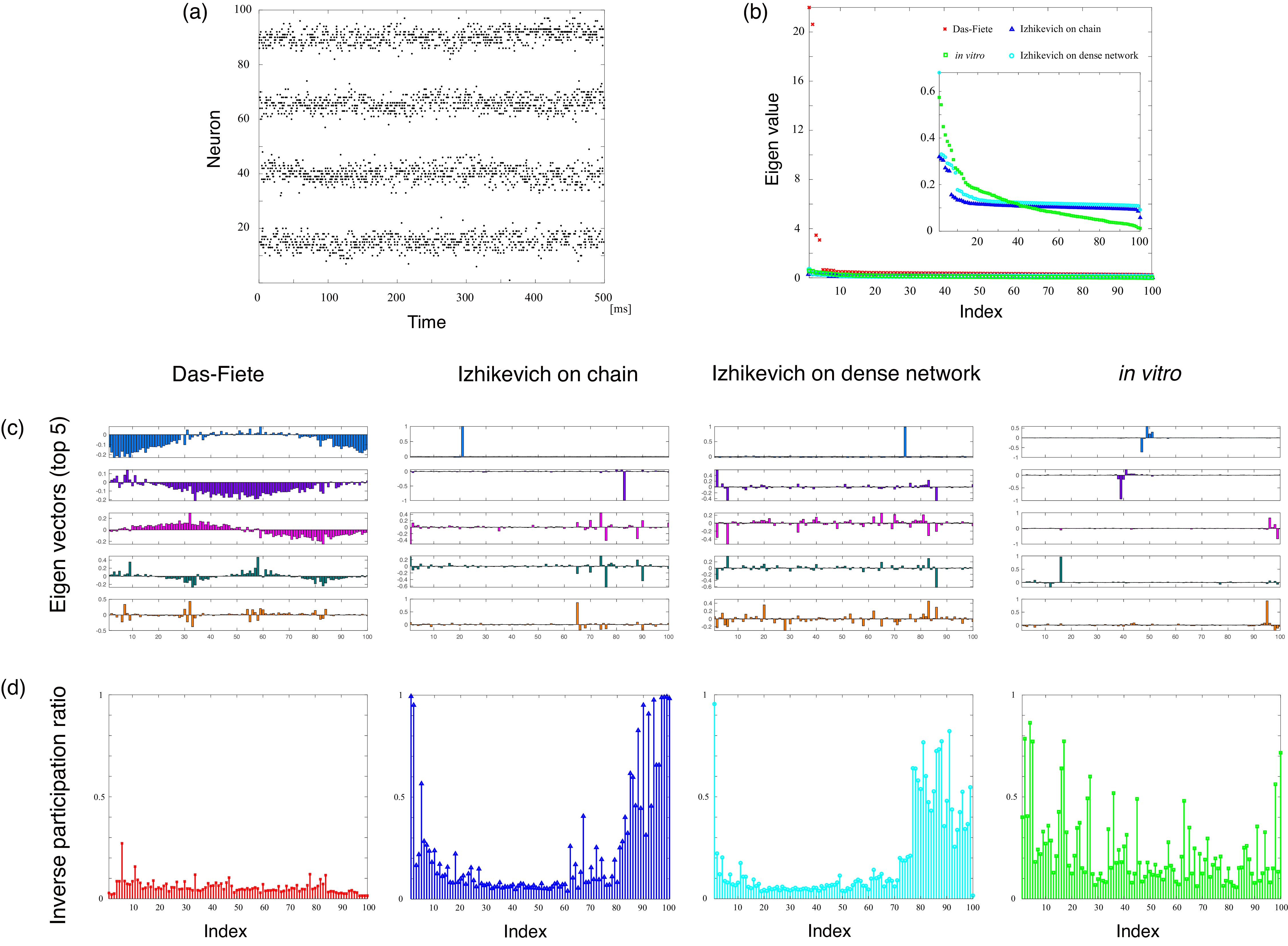}
  \caption{(a) Spike trains in the Das--Fiete system. Coupling strength is assumed \textcolor{black}{to be} large enough for macroscopic order to emerge.
  (b) Eigen spectrums of covariance matrices for the Das--Fiete system, the Izhikevich system on the chain, the Izhikevich system on the dense random network with $q=0.9$,  and the cultured system. 
  \textcolor{black}{The inset} shows the last three cases.
  \textcolor{black}{(c,d)} \textcolor{black}{Eigenvectors} and \textcolor{black}{IPRs from Eq.} \eqref{eq:ipr}.
  \textcolor{black}{(c)} Eigenvectors for the five largest \textcolor{black}{eigenvalues} (the top is the first largest\textcolor{black}{,} the bottom is the fifth largest) and \textcolor{black}{(d) IPRs}\textcolor{black}{, where for the latter the indices are ordered so that the eigenvalues are descending}.}
  \label{fig:eigen_vectors}
\end{figure}

\begin{figure}[h]
  \centering
	\includegraphics[scale=0.35]{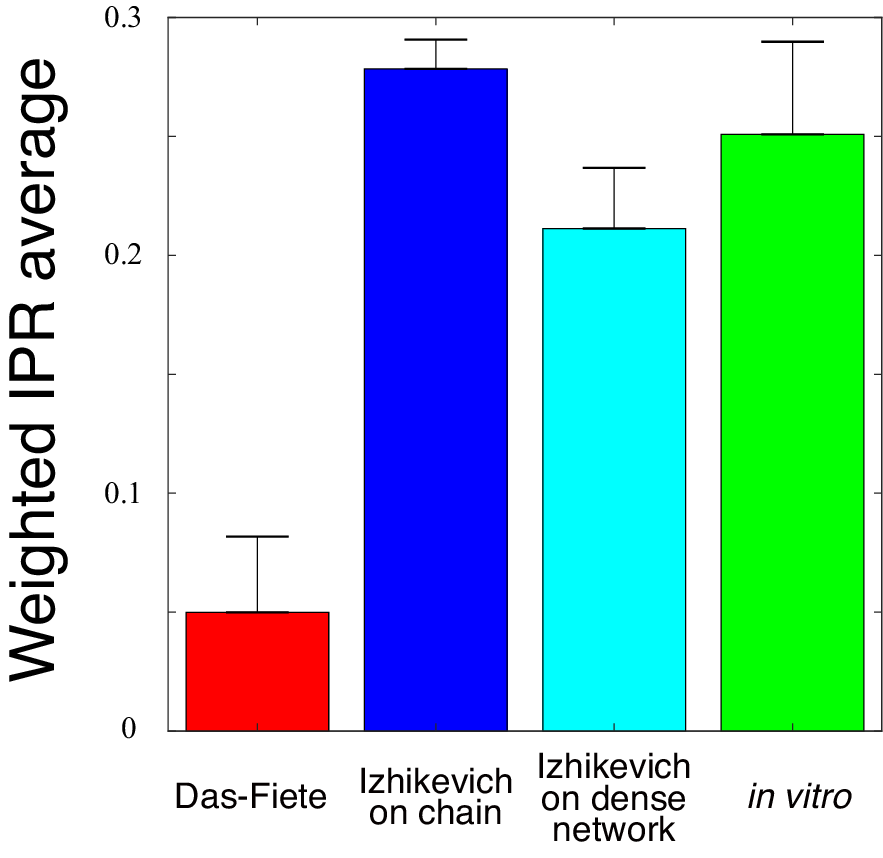}
  \caption{\textcolor{black}{Weighted IPR averages from Eq.} \eqref{eq:weighted_ipr} \textcolor{black}{for the} Das--Fiete system,
  the Izhikevich systems on the chain and the dense network, and the cultured system.
  The error bars indicate the standard \textcolor{black}{deviations} of \textcolor{black}{five} independent simulations for the first three \textcolor{black}{models} and \textcolor{black}{five} different durations for the \textit{in vitro} system.
  }
  \label{fig:weighted_ipr}
\end{figure}

We consider an extreme case with strong couplings \textcolor{black}{that} corresponds to \textcolor{black}{the} $r=0.025$ case in \citep{das2019systematic}.
We set the same model parameters as described in \citep{das2019systematic}, and use $T=2000\,\mathrm{s}$ data averaged over time\textcolor{black}{, to evaluate statistical quantities}.
The spike trains during $500\,\mathrm{ms}$ for the stationary state are shown in Fig. \ref{fig:eigen_vectors} (a).
We observe a macroscopic order in their dynamics, which are induced by Mexican-hat-like inhibitory couplings.
\textcolor{black}{Das and Fiete demonstrated that} this type of dynamics masks the statistics generated by local couplings.
Therefore, we would like to \textcolor{black}{identify these situations} in advance and avoid the possibility of unexpected failure.
To detect possible failure, we look at \textcolor{black}{the eigenvalue} spectrum of the covariance matrix $C_{ij}=\left<s_i(t)s_j(t)\right>-\left<s_i(t)\right>\left<s_j(t)\right>$, \textcolor{black}{also} known as the principal component analysis.
After making the spike trains coarse-grained with the optimal $\Delta\tau_{\mathrm{opt}}$ ($10\,\mathrm{ms}$ for the Das--Fiete system), we \textcolor{black}{calculate the eigenvalues} of the covariance matrices for the Das--Fiete system, the Izhikevich \textcolor{black}{chain system}, the Izhikevich \textcolor{black}{dense random network system} with $q=0.9$, and the cultured system, which are shown in Fig. \ref{fig:eigen_vectors} (b)\textcolor{black}{.}
\textcolor{black}{The data used in the latter three cases are the same data} as above.
In the Das--Fiete system, the maximum eigenvalue (the first principal component) is fairly large, $\sim\mathcal{O}(10)$, whereas the Izhikevich and \textit{in vitro} systems show reasonable $\sim\mathcal{O}(1)$ values. 
The \textcolor{black}{eigenvectors} for the five largest \textcolor{black}{eigenvalues} are shown in Fig. \ref{fig:eigen_vectors} (c).
In the Das--Fiete system we observe long-range modes\textcolor{black}{, whereas they are not present} in the dominant modes of the other three cases.
To characterize the \textcolor{black}{eigenmode profile}, we introduce the inverse participation ratio (IPR), which is defined for each mode $i$ by
\begin{align}
\mathrm{IPR}_i = \frac{\sum_{j=1}^N\left(v_j^{i}\right)^4}{\left(\sum_{j=1}^N\left(v_j^{i}\right)^2\right)^2},\label{eq:ipr}
\end{align}
where $v_j^i$ indicates the $j$th element of the \textcolor{black}{eigenvector} for mode $i$.
The IPR quantifies the profile of the \textcolor{black}{eigenmode}, and its large \textcolor{black}{(or small)} value implies the short \textcolor{black}{(or long)} range mode.
The IPRs for the four cases are shown in Fig. \ref{fig:eigen_vectors} (d).
To discriminate whether dominant modes have long-range \textcolor{black}{features}, we calculate the \textcolor{black}{weighted IPR average as}
\begin{align}
\left<\mathrm{IPR}\right> = \frac{\sum_{i=1}^N\lambda_i\times\mathrm{IPR}_i}{\sum_{i=1}^N\lambda_i},\label{eq:weighted_ipr}
\end{align}
\textcolor{black}{, where $\lambda_i$ is the eigenvalue for mode $i$.}
The result is shown in Fig. \ref{fig:weighted_ipr}.
This implies that Das--Fiete system has the dominant long-range modes characterized by $\left<\mathrm{IPR}\right>$, whereas the others, even with the dense networks, do not.
Therefore, \textcolor{black}{$\left<\mathrm{IPR}\right>$ can be used to} discriminate unfavorable situations for coupling inference induced by the dominant long-range modes.

\section{Conclusion}
In this paper we proposed systematic methods equipped with an objective criterion for processing spike time series data in the inverse problem using statistical inference models.
The first method is to appropriately discretize raw data with the time bins.
This method showed superior effectiveness against the conventional cross-correlation method, as demonstrated in Section \ref{sec:crosscorrelation}.
The second method is for effectively screening couplings obtained as the solution of the inverse problem.
\textcolor{black}{We showed this method can be accelerated using the approximation \citep{terada2018objective} with keeping its inference accuracy as much as the $\ell_1$ regularization in Section \ref{sec:regularization}.}
We showed that these methods work well in simulated and \textit{in vitro} neuronal networks.
These results highlight that the proposed inference procedure, with the use of the kinetic asymmetric Ising model, is quite effective and has the potential to infer actual connectivity, including neuronal excitatory and inhibitory properties.
We stress that the methods described herein for pre- and post-processing data can be used with other types of generalized linear models.

\textcolor{black}{We note that strong long-range modes may hide local statistics and in such cases we cannot reconstruct synaptic connectivity from observed spiking data as reported in \citep{das2019systematic}.
However, as shown in \ref{sec:limitation} we can detect such long-range modes from the statistics of the neuronal activity.
We can use the weighted average of the inverse participation ratios in the correlation matrix to discriminate such conditions, and this method may be useful before applying an inference procedure to neural data.}

\textcolor{black}{
We also remark two potential directions in future works.
First, it is important to develop inference methods under hidden neurons since we often cannot access to the whole relevant population \citep{battistin2017learning}.
Statistical-mechanical analyses have been done in symmetric \citep{hunag2015effects} and asymmetric coupling cases \citep{dunn2013learning,battistin2015belief}.
Thus, it would be interesting to investigate whether our methods are effective in those cases.
Second, although our present treatment ignores the time dependency of the external force, it becomes important in many experimental settings where neurons are exposed to non-stationary input \citep{tyrcha2013effect}.
Hence, it is interesting to generalize our present methods and apply the generalized one to those cases with non-stationary inputs.}

\section*{Acknowledgments}

We thank Abhranil Das and Ila R. Fiete for providing us with a sample code of their work \citep{das2019systematic}.
We are grateful to Yasser Roudi and John Hertz for fruitful discussion.
This work was supported by the Special Postdoctoral Research Program at RIKEN (YT), MEXT KAKENHI Grant Numbers 17H00764 (YT, TO, and YK), 18K11463 (TO), 19H01812 (TO), and 19K20365 (YT), and JST CREST Grant Number JPMJCR1912 (TO and YK).
\textcolor{black}{TO is also supported by a Grant for Basic Science Research Projects from the Sumitomo Foundation.}


\begin{thebibliography}{100}
\providecommand{\natexlab}[1]{#1}
\expandafter\ifx\csname urlstyle\endcsname\relax
  \providecommand{\doi}[1]{doi:\discretionary{}{}{}#1}\else
  \providecommand{\doi}{doi:\discretionary{}{}{}\begingroup
  \urlstyle{rm}\Url}\fi

\bibitem[{Aru et~al.(2015)Aru, Aru, Priesemann, Wibral, Lana, Pipa, Singer, \& Vicente}]{aru2015untangling}
Aru, J., Aru, J., Priesemann, V., Wibral, M., Lana, L., Pipa, G., Singer, W., \& Vicente, R. (2015).
\newblock Untangling cross-frequency coupling in neuroscience.
\newblock \emph{Current Opinion in Neurobiology}, \emph{31}, 51 -- 61.

\bibitem[{Aurell \& Ekeberg(2012)}]{aurell2012inverse}
 Aurell, E. \& Ekeberg, M. (2012).
\newblock Inverse Ising inference using all the data.
\newblock \emph{Physical Review Letters}, \emph{108}, 090201.

\bibitem[{Battistin, Dunn, \& Roudi(2017)}]{battistin2017learning}
\textcolor{black}{Battistin, C., Dunn, B., \& Roudi, Y. (2017).
\newblock Learning with unknowns: Analyzing biological data in the presence of hidden variables.
\newblock \emph{Current Opinion in System Bilogy}, \emph{1}, 122 -- 128.}

\bibitem[{Battistin, Hertz, Tyrcha, \& Roudi(2015)}]{battistin2015belief}
\textcolor{black}{Battistin, C., Hertz, J., Tyrcha, J., \& Roudi, Y. (2015).
\newblock Belief propagation and relicas for inference and learning in a kinetic Ising model with hidden spins.
\newblock \emph{Journal of Statistical Mechanics: Theory and Experiment}, \emph{2015}, P05021.}

\bibitem[{Bialek, W. \& van Steveninck, R. R.(2005)Bialek, \& van Steveninck}]{bialek2005features}
 Bialek, W. \& van Steveninck, R. R. (2005).
\newblock Features and dimensions: Motion estimation in fly vision.
\newblock \emph{arXiv preprint}, q-bio/0505003.

\bibitem[{Brown, Kass, \& Mitra(2004)}]{brown2004multiple} 
Brown, E. N., Kass, R. E., \& Mitra, P. P. (2004).
\newblock Multiple neural spike train data analysis: state-of-the-art and future challenges.
\newblock \emph{Nature Neuroscience}, \emph{7}, 456 -- 461.

\bibitem[{Bulso, Marsili, \& Roudi(2016)}]{bulso2016sparse}
\textcolor{black}{Bulso, N., Marsili, M., \& Roudi, Y. (2016).
\newblock Sparse model selection in the highly under-sampled regime.
\newblock \emph{Journal of Statistical Mechanics: Theory and Experiment}, \emph{2016}(9), 093404.}

\bibitem[{Buzs{\'a}ki(2004)}]{buzsaki2004large} 
Buzs{\'a}ki,, G. (2004).
\newblock Large-scale recording of neuronal ensembles.
\newblock \emph{Nature Neuroscience}, \emph{7}, 446 -- 451.

\bibitem[{Capone, et~al.(2015)Capone Filosa, Gigante, Ricci-Tersenghi, \& Del Giudice}]{capone2015inferring}
\textcolor{black}{Capone, C., Filosa, C., Gigante, G., Ricci-Tersenghi, F., \& Del Giudice, P. (2015).
\newblock Inferring Synaptic Structure in Presence ofNeural Interaction Time Scales.
\newblock \emph{PLoS One}, \emph{10}(3), 0118412.}

\bibitem[{Cheng, et~al.(2011)Cheng, Gon{\c{c}}alves, Golshani, Arisaka, \& Portera-Cailliau}]{cheng2011simultaneous}
Cheng, A., Gon{\c{c}}alves, J. T., Golshani, P., Arisaka, K., \& Portera-Cailliau, C. (2011).
\newblock Simultaneous two-photon calcium imaging at different depths with spatiotemporal multiplexing.
\newblock \emph{Nature Methods}, \emph{8}, 139 -- 142.

\bibitem[{Churchland, Yu, Sahani, \& Shenoy(2007)}]{churchland2007techniques}
 Churchland, M. M., Yu, B. M., Sahani, M. \& Shenoy, K. V. (2007).
\newblock Techniques for extracting single-trial activity patterns from large-scale neural recordings.
\newblock \emph{Current Opinion in Neurobiology}, \emph{17}, 609 -- 618.

\bibitem[{Cocco, Leibler, \& Monasson(2009)}]{cocco2009neuronal}
 Cocco, S., Leibler, S., \& Monasson, R. (2009).
\newblock Neuronal couplings between retinal ganglion cells inferred by efficient inverse statistical physics methods.
\newblock \emph{Proceedings of the National Academy of Sciences}, \emph{106}(33), 14058 -- 14062.

\bibitem[{Cox(1958)}]{cox1958regression}
 Cox, D. R. (1958).
\newblock The regression analysis of binary sequences.
\newblock \emph{Journal of the Royal Statistical Society: Series B (Methodological)}, \emph{20}(2), 215 --242.

\bibitem[{Cunningham \& Yu(2008)}]{cunningham2014dimensionality}
 Cunningham, J. P. \& Yu, B. M. (2008).
\newblock Dimensionality reduction for large-scale neural recordings.
\newblock \emph{Nature Neuroscience}, \emph{17}, 1500 -- 1509.

\bibitem[{Dahmen, Gr{\"u}n, Diesmann, Helias.(2019)}]{dahmen2019second}
 Dahmen D., Gr{\"u}n S.,  Diesmann M., \& Helias M. (2019).
\newblock Second type of criticality in the brain uncovers rich multiple-neuron dynamics.
\newblock \emph{Proceedings of the National Academy of Sciences}, \emph{116}(26), 13051 -- 13060.

\bibitem[{Das \& Fiete(2019)}]{das2019systematic}
 Das A. \& Fiete I. R. (2019).
\newblock Systematic errors in connectivity inferred from activity in strongly coupled recurrent circuits.
\newblock \emph{bioRxiv}, 512053.

\bibitem[{Decelle, \& Zhang(2015)}]{decelle2015inference}
 \textcolor{black}{Decelle, A. \& Zhang, P. (2015).
\newblock Inference of the sparse kinetic Ising model using the decimation method.
\newblock \emph{Physical Review E}, \emph{91}(9), 052136.}

\bibitem[{Dunn, M{\o}rreaunet, \& Roudi(2015)}]{dunn2015correlations}
 Dunn, B., M{\o}rreaunet, M., \& Roudi, Y. (2015).
\newblock Correlations and functional connections in a population of grid cells.
\newblock \emph{PLoS Computational Biology}, \emph{11}(2), e1004052.

\bibitem[{Dunn \& Roudi(2013)}]{dunn2013learning}
\textcolor{black}{Dunn, B. \& Roudi, Y. (2013).
\newblock Learning and inference in a nonequilibrium Ising model with hidden nodes.
\newblock \emph{Physical Review E}, \emph{87}, 022127.}

\bibitem[{Field, et~al.(2010)Field, Gauthier, Sher, Greschner, Machado, Jepson. Shlens, Gunning, Mathieson, Dabrowski, Paninski, Litke, \& Chichilnisky}]{field2010functional}
 \textcolor{black}{Field, G. D., Gauthier, J. L., Sher, A., Greschner, M., Machado, T. A., Jepson, L. H., Shlens, J., Gunning, D. E., Mathieson, K., Dabrowski, W., Paninski, L., Litke, A. M., \& Chichilnisky, E. J. (2010).
\newblock Functional connectivity in the retina at the resolution of photoreceptors.
\newblock \emph{Nature}, \emph{467}, 673 -- 677.}

\bibitem[{Grewe, et~al.(2010)Grewe, Langer, Kasper, Kampa, \& Helmchen}]{grewe2010high}
 Grewe, B. F., Langer, D., Kasper, H., Kampa, B. M. \& Helmchen, F. (2010).
\newblock High-speed in vivo calcium imaging reveals neuronal network activity with near-millisecond precision.
\newblock \emph{Nature Methods}, \emph{7}(5), 399 -- 405.

\bibitem[{Grosmark, \& Buzs{\'a}ki(2016)}]{grosmark2016diversity}
 Grosmark, A. D. \& Buzs{\'a}ki, G., (2016).
\newblock Diversity in neural firing dynamics supports both rigid and learned hippocampal sequences.
\newblock \emph{Science}, \emph{351}(6280), 1440 -- 1443.

\bibitem[{Huang(2015)}]{hunag2015effects}
 \textcolor{black}{Huang, H., (2015).
\newblock Effects of hidden nodes on network structure inference.
\newblock \emph{Journal of Physics A: Mathematical and Theoretical}, \emph{48}, 355002.}

\bibitem[{Ikegaya, et~al.(2004)Ikegaya, Aaron, Cossart, Aronov, Lampl, Ferster, \& Yuste}]{ikegaya2004synfire}
 Ikegaya, Y., Aaron, G., Cossart, R., Aronov, D., Lampl, I., Ferster, D., \& Yuste, R. (2004).
\newblock Synfire chains and cortical songs: temporal modules of cortical activity.
\newblock \emph{Science}, \emph{559}(5670), 559 -- 564.

\bibitem[{Isomura, et~al.(2015)Isomura, Shimba, Takayama, Takeuchi, Kotani, \& Jimbo}]{isomura2015signal}
 Isomura, T., Shimba, K., Takayama, Y., Takeuchi, A., Kotani, K., \& Jimbo, Y. (2015).
\newblock Signal transfer within a cultured asymmetric cortical neuron circuit.
\newblock \emph{Isomura, Takuya, et al. "Signal transfer within a cultured asymmetric cortical neuron circuit." Journal of neural engineering}, \emph{12}(6), 066023.

\bibitem[{Izhikevich(2003)}]{izhikevich2003simple}
 Izhikevich, E. M. (2003).
\newblock Simple model of spiking neurons.
\newblock \emph{IEEE Transactions on neural networks}, \emph{14}(6), 1569 -- 1572.

\bibitem[{Kappen \& Rodr{\'\i}guez(1998)}]{kappen1998efficient}
 Kappen, H. J. \& Rodr{\'\i}guez, F. d. B. (1998).
\newblock Efficient learning in Boltzmann machines using linear response theory.
\newblock \emph{Neural Computation}, \emph{10}(5), 1137 -- 1156.

\bibitem[{Kobayashi, et~al.(2019)Kobayashi, Kurita, Kurth, Kitano, Mizuseki, Diesmann, Richmond \& Shinomoto}]{kobayashi2019reconstructing}
 \textcolor{black}{Kobayashi, R., Kurita, S., Kurth, A., Kitano, K., Mizuseki, K., Diesmann, M., Richmond, B. J., \& Shinomoto, S. (2019).
\newblock Reconstructing neuronal circuitry from parallel spike trains.
\newblock \emph{Nature Communications}, \emph{10}, 4468.}

\bibitem[{Kullback(1997)}]{kullback1997information}
 Kullback, S. (1997).
\newblock Information theory and statistics.
\newblock Dover, New York.

\bibitem[{Lezon, et~al.(2006)Lezon, Banavar, Cieplak, Maritan, \& Fedoroff}]{lezon2006using}
 Lezon, T. R., Banavar, J. R., Cieplak, M., Maritan, A., \& Fedoroff, N. V. (2006).
\newblock Using the principle of entropy maximization to infer genetic interaction networks from gene expression patterns.
\newblock \emph{Proceedings of the National Academy of Sciences}, \emph{103}(50), 19033 -- 19038.

\bibitem[{Maass(2016)}]{maass2016searching}
 Maass, W. (2016).
\newblock Searching for principles of brain computations.
\newblock \emph{Current Opinion in Behavioral Sciences}, \emph{11}, 81 -- 92.

\bibitem[{Marom \& Shahaf(2002)}]{marom2002development}
 \textcolor{black}{
 Marom, S. \& Shahaf, G. (2002).
\newblock Development, learning and memory in large random networks of cortical neurons: lessons beyond anatomy.
\newblock \emph{Quarterly reviews of biophysics}, \emph{35}(1), 63 -- 87.}

\bibitem[{McCulloch \& Pitts(1943)}]{mcculloch1943logical}
 McCulloch, W. S. \& Pitts, W. (1943).
\newblock A logical calculus of the ideas immanent in nervous activity.
\newblock \emph{The bulletin of mathematical biophysics}, \emph{5}(4), 115 -- 133.

\bibitem[{M{\'e}zard \& Sakellariou(2011)}]{mezard2011exact}
 M{\'e}zard, M. \& Sakellariou, J. (2011).
\newblock Exact mean-field inference in asymmetric kinetic Ising systems.
\newblock \emph{Journal of Statistical Mechanics: Theory and Experiment}, \emph{2011}(07), L07001.

\bibitem[{Ohiorhenuan, et~al.(2010)Ohiorhenuan, Mechler, Purpura, Schmid, Hu, \& Victor}]{ohiorhenuan2010sparse}
 Ohiorhenuan, I. E., Mechler, F., Purpura, K. P., Schmid, A. M., Hu, Q. \& Victor, J. D. (2010).
\newblock Sparse coding and high-order correlations in fine-scale cortical networks.
\newblock \emph{Nature}, \emph{466}(7306), 617.

\bibitem[{Okatan, Wilson, \& Brown(2005)}]{okatan2005analyzing}
 \textcolor{black}{Okatan, M., Wilson, M. A., \& Brown, E. N. (2005).
\newblock Analyzing Functional Connectivity Using a NetworkLikelihood Model of Ensemble Neural Spiking Activity.
\newblock \emph{Neural Computation}, \emph{17}, 1927 -- 1961.}

\bibitem[{Ostojic, Brunel, \& Hakim(2009)}]{ostojic2009how}
 \textcolor{black}{Ostojic, S., Brunel, N., \& Hakim, V. (2009).
\newblock How connectivity, background activity, and synaptic properties shape the cross-correlation between spike trains.
\newblock \emph{Journal of Neuroscience}, \emph{29}(33), 10234 -- 10253.}

\bibitem[{Paninski \& Cunningham(2018)}]{paninski2018neural}
 Paninski, L. \& Cunningham, J. P. (2018).
\newblock Neural data science: accelerating the experiment-analysis-theory cycle in large-scale neurosciences.
\newblock \emph{Current Opinion in Neurobiology}, \emph{50}, 232 -- 241.

\bibitem[{Pillow, et~al.(2008)Pillow, Shlens, Paninski, Sher, Litke, Chichilnisky \& Simoncelli}]{pillow2008spatio}
 \textcolor{black}{Pillow, J. W., Shlens, J., Paninski, L., Sher, A., Litke, A. M., Chichilnisky, E. J., \& Simoncelli, E. P. (2008).
\newblock Spatio-temporal correlations and visual signalling in a complete neuronal population.
\newblock \emph{Nature}, \emph{454}, 995 -- 999.}

\bibitem[{Roudi, Dunn, \& Hertz(2015)}]{roudi2015multi}
 Roudi, Y., Dunn, B., \& Hertz, J. (2015).
\newblock Multi-neuronal activity and functional connectivity in cell assemblies.
\newblock \emph{Current Opinion in Neurobiology}, \emph{32}, 38 -- 44.

\bibitem[{Roudi \& Hertz(2011)}]{roudi2011mean}
 Roudi, Y. \& Hertz, Y. (2011).
\newblock Mean field theory for nonequilibrium network reconstruction.
\newblock \emph{Physical Review Letters}, \emph{106}, 048702.

\bibitem[{Roudi, Tyrcha, \& Hertz(2009)}]{roudi2009ising}
 Roudi, Y., Tyrcha, J., \& Hertz, J. (2009).
\newblock Ising model for neural data: model quality and approximate methods for extracting functional connectivity.
\newblock \emph{Physical Review E}, \emph{79}, 051915.

\bibitem[{Sakellariou, Roudi, Mezard, \& Hertz(2012)}]{sakellariou2012effect}
 Sakellariou, J., Roudi, Y., Mezard, M., \& Hertz, J. (2012).
\newblock Effect of coupling asymmetry on mean-field solutions of the direct and inverse Sherrington--Kirkpatrick model.
\newblock \emph{Philosophical Magazine}, \emph{92}(1-3), 272 -- 279.

\bibitem[{Schneidman, Berry, Segev, \& Bialek(2006)}]{schneidman2006weak}
 Schneidman, E., Berry, M. J., Segev, R., \& Bialek, W. (2006).
\newblock Weak pairwise correlations imply strongly correlated network states in a neural population.
\newblock \emph{Nature}, \emph{440}(7087), 1007 -- 1012.

\bibitem[{Schwartz, Pillow, Rust, \& Simoncelli(2006)}]{schwartz2006spike}
 Schwartz, O., Pillow, J. W., Rust, N. C., \& Simoncelli, E. P. (2006).
\newblock Spike-triggered neural characterization.
\newblock \emph{Journal of vision}, \emph{6}(4), 13 -- 13.

\bibitem[{Sessak, \& Monasson(2009)}]{sessak2009small}
 Sessak, V. \& Monasson, R. (2009).
\newblock Small-correlation expansions for the inverse Ising problem.
\newblock \emph{Journal of Physics A: Mathematical and Theoretical}, \emph{42}(5), 055001.

\bibitem[{Shlens, et~al.(2006)Shlens, Field, Gauthier, Grivich, Petrusca, Sher, Litke, \& Chichilnisky}]{shlens2006structure}
 Shlens, J., Field, G. D., Gauthier, J. L., Grivich, M. I., Petrusca, D., Sher, A., Litke, A. M., \& Chichilnisky, E. J. (2006).
\newblock The structure of multi-neuron firing patterns in primate retina.
\newblock \emph{Journal of Neuroscience}, \emph{26}(32), 8254 -- 8266.

\bibitem[{Sokalr \& Rohlf(2011)}]{sokalr1981principles}
 \textcolor{black}{Sokal, R. R. \& Rohlf, F. J. (2011).
\newblock Biometry: The principles and practice of statistics in biological research.
\newblock \emph{W. Ho Freeman}, San Francisco.}

\bibitem[{Steinmetz, et~al.(2018)Steinmetz, Koch, Harris, \& Carandini}]{steinmetz2018challenges}
 \textcolor{black}{Steinmetz, N. A., Koch, C, Harris, K D., \& Carandini, M. (2018).
\newblock Challenges and opportunities for large-scale electrophysiology with Neuropixels probes.
\newblock \emph{Current Opinion in Neurobiology}, \emph{50}, 92 -- 100.}

\bibitem[{Stringer, et~al., (2019)Stringer, Pachitariu, Steinmetz, Carandini, \& Harris}]{stringer2019high}
 Stringer, C., Pachitariu, M., Steinmetz, N., Carandini, M. \& Harris, K. D. (2019).
\newblock High-dimensional geometry of population responses in visual cortex.
\newblock \emph{Nature}, \emph{571}, 361 -- 365.

\bibitem[{Takekawa, Isomura, \& Fukai(2010)}]{takekawa2010accurate}
 Takekawa, T., Isomura, Y., \& Fukai, T. (2010).
\newblock Accurate spike sorting for multi-unit recordings.
\newblock \emph{European Journal of Neuroscience}, \emph{31}(2), 263 -- 272.

\bibitem[{Tanaka(1998)}]{tanaka1998mean}
 Tanaka, T. (1998).
\newblock Mean-field theory of Boltzmann machine learning.
\newblock \emph{Physical Review E}, \emph{58}, 2302.

\bibitem[{Tang, et~al.(2008)}]{tang2008maximum}
 Tang, A., Jackson, D., Hobbs, J., Chen, W., Smith, J. L., Patel, H., Prieto, A., Petrusca, D., Grivich, M. I., Sher, A., Hottowy, P., Dabrowski, W., Litke A. M., \& Beggs, J. M. (2008).
\newblock A maximum entropy model applied to spatial and temporal correlations from cortical networks in vitro.
\newblock \emph{Journal of Neuroscience}, \emph{28}(2), 505 -- 518.

\bibitem[{Terada, Obuchi, Isomura, \& Kabashima(2018)}]{terada2018objective}
 Terada, Y., Obuchi, T., Isomura, T., \& Kabashima, Y. (2018).
\newblock Objective and efficient inference for couplings in neuronal networks.
\newblock \emph{Advances in Neural Information Processing Systems}, 4971 -- 4980.

\bibitem[{Tyrcha, Roudi, Marsili, \& Hertz(2013)}]{tyrcha2013effect}
 Tyrcha, J., Roudi, Y., Marsili, M., \& Hertz, J. (2013).
\newblock The effect of nonstationarity on models inferred from neural data.
\newblock \emph{Journal of Statistical Mechanics: Theory and Experiment}, \emph{2013}(03), P03005.

\bibitem[{Weigt, et~al.(2009)Weigt, White, Szurmant, Hoch, \& Hwa}]{weigt2009identification}
 Weigt, M., White, R. A., Szurmant, H., Hoch, J. A., \& Hwa, T. (2009).
\newblock Identification of direct residue contacts in protein--protein interaction by message passing.
\newblock \emph{Proceedings of the National Academy of Sciences}, \emph{106}(1), 67 -- 72.

\bibitem[{Xu, Aurell, Corander, \& Kabashima(2017)}]{xu2017statistical}
 Xu, Y., Aurell, E., Corander, J., \& Kabashima, Y. (2017).
\newblock Statistical properties of interaction parameter estimates in direct coupling analysis.
\newblock \emph{arXiv}, 1704.01459.

\bibitem[{Xu, Puranen, Corander, \& Kabashima(2018)}]{xu2018inverse}
\textcolor{black}{Xu, Y., Puranen, S., Corander, J., \& Kabashima, Y. (2018).
\newblock Inverse finite-size scaling for high-dimensional significance analysis.
\newblock \emph{Physical Review E}, \emph{97}, 062112}

\bibitem[{Yuste(2015)}]{yuste2015neuron}
 Yuste, R. (2015).
\newblock From the neuron doctrine to neural networks.
\newblock \emph{Nature Reviews Neuroscience}, \emph{16}(8), 487 -- 497.

\bibitem[{Zeng, et~al.(2013)Zeng, Alava, Aurell, Hertz, \& Roudi}]{zeng2013maximum}
 Zeng, H. L., Alava, M, Aurell, E, Hertz, J., \& Roudi, Y. (2013).
\newblock Maximum likelihood reconstruction for Ising models with asynchronous updates.
\newblock \emph{Physical Review Letters}, \emph{110}, 210601.

\bibitem[{Zeng, Aurell, Alava, \& Mahmoudi(2011)}]{zeng2011network}
 Zeng, H. L., Aurell, E., Alava, M., \& Mahmoudi, H. (2011).
\newblock Network inference using asynchronously updated kinetic Ising model.
\newblock \emph{Physical Review E}, \emph{83}, 041135.

\bibitem[{Zeng, Hertz, \& Roudi(2014)}]{zeng2014l1}
 \textcolor{black}{Zeng, H. L., Hertz, J., \& Roudi, Y. (2014).
\newblock L1 regularization for reconstruction of a non-equilibrium Ising model.
\newblock \emph{Physics Scripta}, \emph{89}, 105002.}


\end{thebibliography}
%
%
%

\bibliographystyle{APA}








\end{document}